\documentclass[twocolumn,showpacs,prl,footinbib,superscriptaddress,amsmath,amssymb,floatfix]{revtex4-2}

\usepackage{graphicx,color}
\usepackage{dcolumn}
\usepackage{bm}
\usepackage{lineno, soul, ulem}
\usepackage{hyperref,url}
\usepackage{xcolor}
\usepackage{array}
\usepackage{ulem}

\newcommand{\healpix}[1]{\texttt{HEALPix} #1}
\newcommand{\emcee}[1]{\texttt{emcee} #1}

 \hypersetup{
    colorlinks=true,
    linkcolor=blue,
    filecolor=blue,      
    urlcolor=blue,
    citecolor=blue,
    pdftitle={Power-spectrum space decomposition of frequency tomographic data for intensity mapping experiments}
    }

\newcommand{\apjl}{Astrophys J.}
\newcommand{\apjs}{Astrophys. J. Suppl. Ser.}
\newcommand{\aap}{Astron. Astrophys.}

\newcommand{\jcap}{J. Cosmol. Astropart. Phys.}
\newcommand{\mnras}{Mon. Not. R. Astron. Soc.}

\newcommand{\pasp}{Publications of the Astronomical Society of the Pacific}

\newcommand{\HI}{21-cm }

\newcommand{\psdfttitle}{Power-spectrum space decomposition of frequency tomographic data for intensity mapping experiments}

\begin{document}

\title{\psdfttitle}

\author{Chang Feng}
\email{changfeng@ustc.edu.cn (corresponding author)}
\affiliation{Department of Astronomy, School of Physical Sciences, University of Science and Technology of China, Hefei, Anhui 230026, China}
\affiliation{CAS Key Laboratory for Research in Galaxies and Cosmology, University of Science and Technology of China, Hefei, Anhui 230026, China}
\affiliation{School of Astronomy and Space Science, University of Science and Technology of China, Hefei, Anhui 230026, China}

 \author{Filipe B. Abdalla}
\affiliation{University College London, Gower Street, London, WC1E 6BT, UK}
\affiliation{Instituto de F\'{i}sica, Universidade de S\~ao Paulo, R. do Mat\~ao, 1371 - Butant\~a, 05508-09 - S\~ao Paulo, SP, Brazil}
\affiliation{Department of Physics and Electronics, Rhodes University, PO Box 94, Grahamstown, 6140, South Africa}

\begin{abstract}
We present a Bayesian framework to establish a power-spectrum space decomposition of frequency tomographic (PSDFT) data for future intensity mapping (IM) experiments. Different from most traditional component-separation methods which work in the map domain, this new technique treats multifrequency power spectra as raw data and can reconstruct component power spectra by taking advantage of distinct components' correlation patterns in the frequency domain. We have validated this new technique for both interferometric and single-dish-like IM experiments, respectively, using synthesized mock data that contain bright foreground contaminants, IM signals, and instrumental effects at different frequencies. The PSDFT approach can effectively remove the bright foreground contamination and extract the targeted IM signals using a Bayesian approach in a power-spectrum subspace. This new approach can be directly applied to a broad range of IM analyses and will be well suited to future high-quality IM datasets, providing a powerful tool for future IM surveys.
\end{abstract}

\maketitle

\textit{Introduction.---}\label{intro}
Spin-flip transitions of neutral hydrogen (HI) atoms, which are the simplest and most abundant form of baryons, can emit photons at \HI wavelengths. These neutral hydrogen atoms are spatially distributed and have peculiar motions, thereby the \HI emissions are not only redshifted but also provide us with a unique tracing of the underlying dark matter fluctuations which are crucial for questions in fundamental physics including primordial non-Gaussianity~\cite{2011PhRvL.107m1304J,2013PhRvL.111q1302C}, modified gravity theories~\cite{2010PhRvD..81f2001M,2013PhRvD..87f4026H}, and exotic dark matter models including warm dark matter~\cite{PhysRevD.71.063534,2017PhRvD..96b3522I}, fuzzy dark matter~\cite{2017PhRvL.119c1302I} and axion-like particles~\cite{2021MNRAS.500.3162B}. More specifically, the \HI signature contains tomographic information for a few unique stages in the evolution of our Universe such as the dark ages~\cite{2007PhRvD..76h3005L,2023JApA...44...10B}, the epoch of reionization~\cite{2003ApJ...596....1C}, and the post-reionization eras~\cite{2015ApJ...803...21B}. Moreover, photon-baryon oscillations in the pre-recombination plasma induce ripples on the neutral hydrogen distributions in the post-recombination era, and these ripples are referred to as baryon acoustic oscillations (BAO). The BAO signature is a standard ruler in both modern cosmology and astrophysics and is a sensitive probe to dark energy~\cite{2008PhRvL.100i1303C,2008MNRAS.383.1195W} which leaves unique imprints in the \HI fluctuations.  

Intensity mapping (IM) with atomic lines, particularly neutral hydrogen, can efficiently survey large volumes of the sky at multifrequency radio wavelengths \cite{2015arXiv150104076M, Santos:2015}. However, radio signals originating from our galaxy are a few orders of magnitude stronger than the faint \HI fluctuations, making measurements of the \HI signals extremely difficult. 

So far, evidence for the \HI signatures has been found by cross-correlating the neutral hydrogen observations with galaxy surveys~\cite{2010Natur.466..463C,2022MNRAS.510.3495W,2023ApJ...947...16A}, but the \HI signals have not been directly detected from the neutral hydrogen IM data and only upper bounds have been obtained~\cite{2019ApJ...884....1B,2022ApJ...925..221A,2023arXiv230207969K,2023MNRAS.521.5120K}. The BAO signals have been detected from the cosmic microwave background (CMB)~\cite{2020A&A...641A...5P}, galaxy surveys~\cite{2005ApJ...633..560E} and Lyman-$\alpha$ forest~\cite{2014JCAP...05..027F,2015A&A...574A..59D}, but there is still no evidence of BAO at radio wavelengths. 

Different approaches have been developed to eliminate radio foreground emissions from the raw multifrequency data in map space, such as principal component analysis (PCA)~\cite{2013ApJ...763L..20M} and the internal linear combination (ILC) methods which make use of smooth frequency characteristics of foreground contaminants~\cite{2007astro.ph..2198D,2012MNRAS.419.1163B,2013MNRAS.435...18B}; the independent component analysis (ICA) and Fast ICA methods which make use of non-Gaussianity of raw data~\cite{2012MNRAS.423.2518C, 2013MNRAS.429..165C}; the generalized morphological component analysis (GMCA) which makes use of generalized morphological components of raw data~\cite{2013MNRAS.429..165C}; and the generalized needlet ILC and singular vector
projection (SVP) methods which adopt a priori information of covariance~\cite{2016MNRAS.456.2749O,2023ApJ...945...38Z} on the desired signals.
However, detection of the \HI and BAO signatures with the map-space approaches can not be directly carried out until the strong foreground contamination is largely removed. Future IM surveys will produce multifrequency datasets with large sky coverage, high resolution, and high sensitivity, requiring an arguably prohibitive and definitely undesirable amount of map manipulations if map-based approaches are adopted.

In this work, we introduce a power-spectrum space decomposition of frequency tomographic data (PSDFT) to simultaneously extract radio foreground emissions, neutral hydrogen fluctuations, and the BAO signals from the raw multifrequency data as a proof-of-concept study. This new technique can be directly applied to a broad range of IM analyses and will be well suited to future high-quality IM datasets, opening up a new window on the component separation theories. 

\textit{Intensity mapping signal model.---}\label{modeling}The spatial distribution of neutral hydrogen is a tracer of the underlying dark matter distribution~\cite{2004ApJ...613...16F} so neutral hydrogen density $\delta_{\rm HI}$ can be approximated by dark matter density $\delta_c$ on linear scales thus neutral hydrogen power spectrum $P_{\rm HI}(k)$ is proportional to dark matter power spectrum $P(k)$, i.e., $P_{\rm HI}(k)\sim P(k)$. Here $k$ denotes the spatial and temporal coordinates in the Fourier domain. We note here that this approximation does not include redshift space distortions, however, these terms should not change the power spectrum or the maps significantly, thus we neglect these terms in this work to simplify the modeling of the HI fluctuations and will validate that this approach works well without requiring a particular model.

Radio telescopes measure these fluctuations with narrow bands that integrate neutral hydrogen density in finite redshift ranges. Therefore, the correlation function (or the power spectrum) of fluctuations measured at two neighboring narrow bands $\nu_A$ and $\nu_B$ contains neutral hydrogen signatures in the overlapping redshift window which is a product of redshift windows $W_{\nu_A}$ and $W_{\nu_B}$ determined by $\nu_A$ and $\nu_B$, respectively. Specifically, we can express the angular power spectrum as $C^{\nu_A\nu_B}_{\ell}=\int d\chi/\chi^2W_{\nu_A}(z)W_{\nu_B}(z)P_{\rm HI}(k)$, where $\chi$ is the comoving distance and $\ell$ is the multipole. 

To explore the detectability of the angular power spectra, certain parametrization schemes are required. For example, the angular power spectra $C^{\nu_A\nu_B}_{\ell}$ can be written as geometric products $\sqrt{A_{\nu_A}A_{\nu_B}C^{\nu_A\nu_A}_{\ell}C^{\nu_B\nu_B}_{\ell}}$. However, only global amplitudes $A_{\nu_A}$ and $A_{\nu_B}$ can be inferred from such a parametrization, and the amplitudes have to be positive. Also, the number of free parameters will increase as the number of frequencies increases. To avoid these issues, we adopt a $k$-space wavelet decomposition which applies a $k$-space wavelet window $\eta_i(k)$ to the HI power spectrum 
\begin{eqnarray}
C^{(i),\nu_A\times\nu_B}_{\ell}&\sim&\int \frac{d\chi}{\chi^2}W_{\nu_A}(z)W_{\nu_B}(z)\eta_i(k)P(k),
\end{eqnarray}
so the HI power spectrum can be decomposed into $N_k$ $k$ modes, i.e., $\sum_{i=1}^{N_k}A^{(i)}_kC^{(i),\nu_A\times\nu_B}_{\ell}$. The $k$-space wavelet windows are normalized and $\sum_{i=1}^{{N_k}}\eta_i(k)=1$ where $\eta_i(k)$ is a top-hat window at the $i$-th $k$-mode. 

This parametrization scheme is advantageous in two aspects. The HI power spectra contain $N_k$ free parameters $\{A^{(i)}_k\}$ which do not increase as the number of frequency channels increases, so each $ k$ mode can be constrained with high signal-to-noise ratios. The amplitude of each $k$ wavelet can vary in both positive and negative regimes so a uniform amplitude prior can be applied.  

We generate matter power spectra with ($P^{\rm w}(k)$) and without ($P^{\rm nw}(k)$)  BAO wiggles using the fitting formula in~\cite{1998ApJ...496..605E}. The matter power spectrum with BAO is related to the one without BAO by a ratio $r^{\rm w}(k)=P^{\rm w}(k)/P^{\rm nw}(k)=1+f(k)$ where the function $f(k)$ can encode any new physics. In this work, we adopt the following phenomenological model
\begin{equation}
f(k)=A_{\rm osc}ke^{-(\frac{k}{k_0})^{n_d}}\sin{(2\pi\frac{k}{k_A})}\label{baomodel}
\end{equation} to exclusively account for the BAO signatures~\cite{2003ApJ...594..665B, 2015ApJ...803...21B, 2016MNRAS.460.4210G, 2021MNRAS.506.2638K}. Here the pivot scale is $k_0$, the sound horizon scale is $k_A$, the spectral index is $n_d$ and the overall amplitude is $A_{\rm osc}$. A shift parameter $\alpha$ can be incorporated into the $f(k)$ function as $f(k^{\prime})=f(k/\alpha)$ and is assumed to be a small deviation from unity. We Taylor expand $f(k^{\prime})$ around $\alpha_0$, i.e.,
\begin{equation}
f(k;\alpha)=f(k;\alpha_0)+\displaystyle\sum_{n=1}^{N_{\alpha}}\frac{1}{n!}\frac{d^{(n)}f}{d\alpha^{(n)}}{\Big |}_{\alpha=\alpha_0}(\alpha-\alpha_0)^n.
\end{equation}
Then the BAO-induced power spectra at each multipole band $b$ can be decomposed into different perturbation terms $\Delta^{\nu_A\times\nu_B,{\rm BAO},(n)}_b$ as $C^{\nu_A\times\nu_B,{\rm BAO},(0)}_{b}+\sum_{n=1}^{N_{\alpha}}(\alpha-\alpha_0)^n/{n!}\Delta^{\nu_A\times\nu_B,{\rm BAO},(n)}_b$~\cite{2023MNRAS.519..799H}. Here the superscript ``(n)'' refers to the $n$-th order term [see Eq. (\ref{baopert}) in \textit{Supplemental Material}] and the default shift parameter $\alpha_0=1$.

The model in Eq. (\ref{baomodel}) can be applied to any study of new physics that has an effect measurable by the power spectrum, such as signatures of primordial features~\cite{2016JCAP...09..023C,2023PhRvD.107d3532B}, cold dark matter isocurvature mode~\cite{2022PhRvD.105h3523M}, light thermal relics~\cite{2018JCAP...08..029B}, warm dark matter~\cite{2001ApJ...556...93B}, and ultra-light axion-like particles~\cite{2021JCAP...01..061K,2022JCAP...08..066K}. The phenomenological BAO model used in Eq. (\ref{baomodel}) is only one of the examples.

\textit{Mock power spectra.---}We apply the Cholesky decomposition to the HI covariance matrix in order to generate correlated HI Gaussian realizations. We have verified that the HI realizations have the same power spectra as the input and are correlated as predicted. For radio foreground simulations, we generate synchrotron emission, free-free emission, anomalous microwave emission (AME), and the cosmic microwave background (CMB) using the Python Sky Model~\cite{2017MNRAS.469.2821T}.  

We select an observing region in the southern hemisphere with the Galactic plane excluded and smooth the edges with a $5^{\circ}$ taper to mitigate sky signal mixing at the Fourier space. The masked sky region has a 35\% sky fraction which is at a similar coverage level to many ongoing ground-based IM experiments. We simulate foreground contaminants, HI, and BAO components from 925 to 1075 MHz with 30 narrow bands. We calculate the power spectra of the masked mock data and bin all the angular power spectra at $50<\ell<500$ with $N_b$ broad bands. We note that the ``broad band'' denotes the binned angular power spectrum and the ``narrow band'' denotes a frequency band with a finite bandwidth $\Delta\nu$ in this work. The foreground contaminants are a few orders of magnitude brighter than the HI signals, meanwhile, the HI signals are about two orders of magnitude stronger than the BAO. 

We assume a constant thermal noise level $N_{\ell}$ among all frequencies and a frequency dependent Gaussian beam profile $b_{\ell}(\nu)=\exp{[-\ell(\ell+1)\theta^2/(16\ln2)]}$ with a full-width-at-half-maximum (FWHM) $\theta(\nu)=\theta_{\rm ref}(\nu_{\rm ref}/\nu)$ where $\theta_{\rm ref}$ is the FWHM at a reference frequency $\nu_{\rm ref}$. These beam profiles are incorporated into the mock data and are eliminated when the mock power spectra are obtained.  In addition, we also consider broadband (\textit{bb}) contributions with polynomials $C^{\nu_A\times\nu_B, {\rm bb},(i)}_{\ell}=\epsilon_i\ell^{i}$ for possible systematic effects~\cite{2012MNRAS.427.2132P}. The default parameters $\{\epsilon_i\}$ for the systematic effects are set to be zero. We note that other possible contaminants may not have smooth frequency spectra, such as polarization leakages. We defer the investigations that differ from the above prescription, including polarization leakage modeling to future work.

\begin{figure}
\includegraphics[width=9cm,height=8cm,trim=100 0 0 0 ]{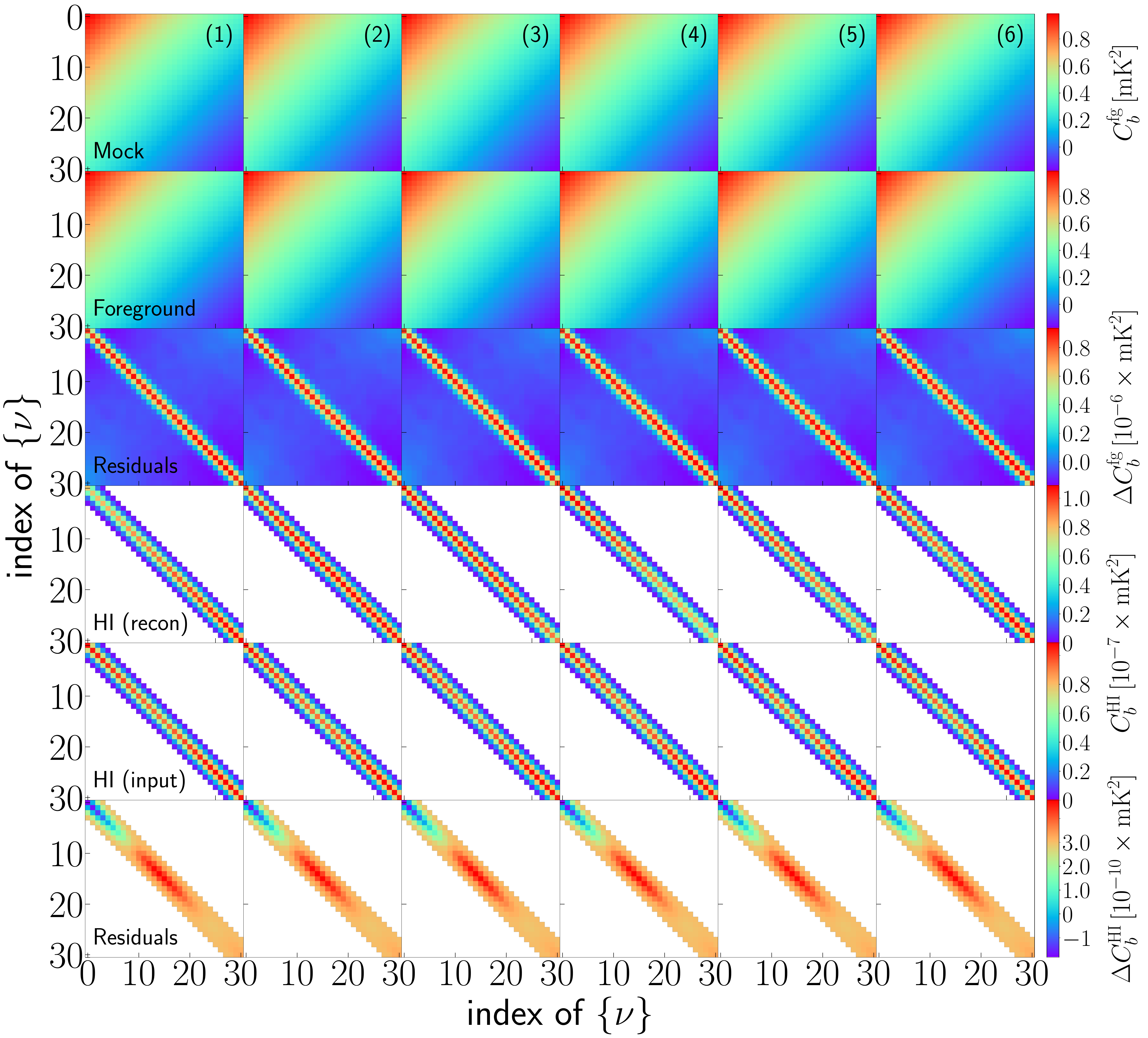}
\caption{Band powers of mock data at all frequencies arranged in a square for each broadband $b$ labeled by ``$(i)$''. The three rows at the top are the raw band powers of the mock data, the foreground models generated from the principal component analysis, and the difference between the first two rows, respectively. The three rows at the bottom are the reconstructed HI signals, input HI signals, and the difference between these two rows in the auto-power spectrum subspace, respectively. From the third row, it is seen that the HI signals become dominant after the foreground subtraction. The axis is arranged by the indices of all the frequencies $\{\nu\}$, and the colorbars generated from the last broadband refer to different band-power values at specific broadbands.}\label{fgfit}
\end{figure}

\begin{figure*}
\includegraphics[width=15.5cm, height=6cm]{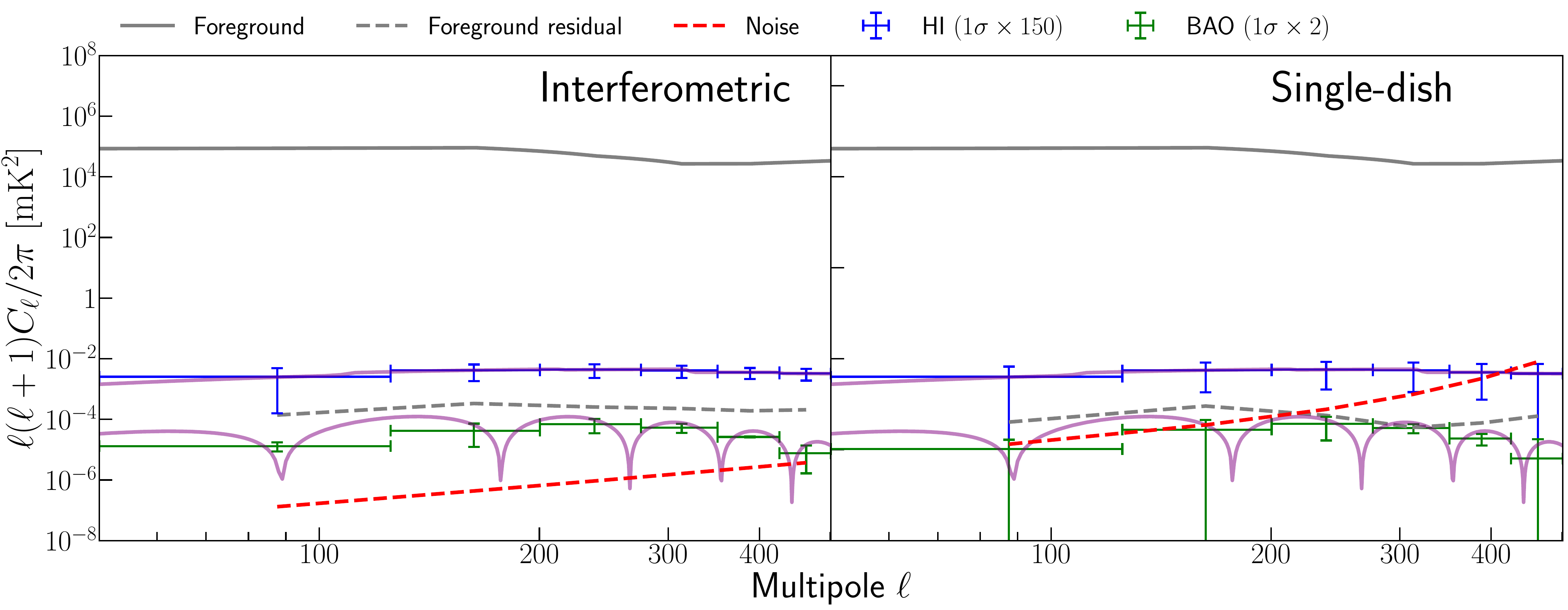}
\caption{Reconstructed band powers of HI (without BAO) and BAO signals at $\nu=980$ MHz. Band powers of HI and BAO signals can be correctly reconstructed from the Bayesian analysis in Eq. (\ref{MLE}) for both the interferometric (left) and single-dish-like IM experiments (right), respectively. Despite orders of magnitude differences in signal strengths, all the reconstructed band powers are consistent with theoretical expectations.}\label{kpdfs2}
\end{figure*}

\begin{figure*}
\includegraphics[width=16cm, height=6cm]{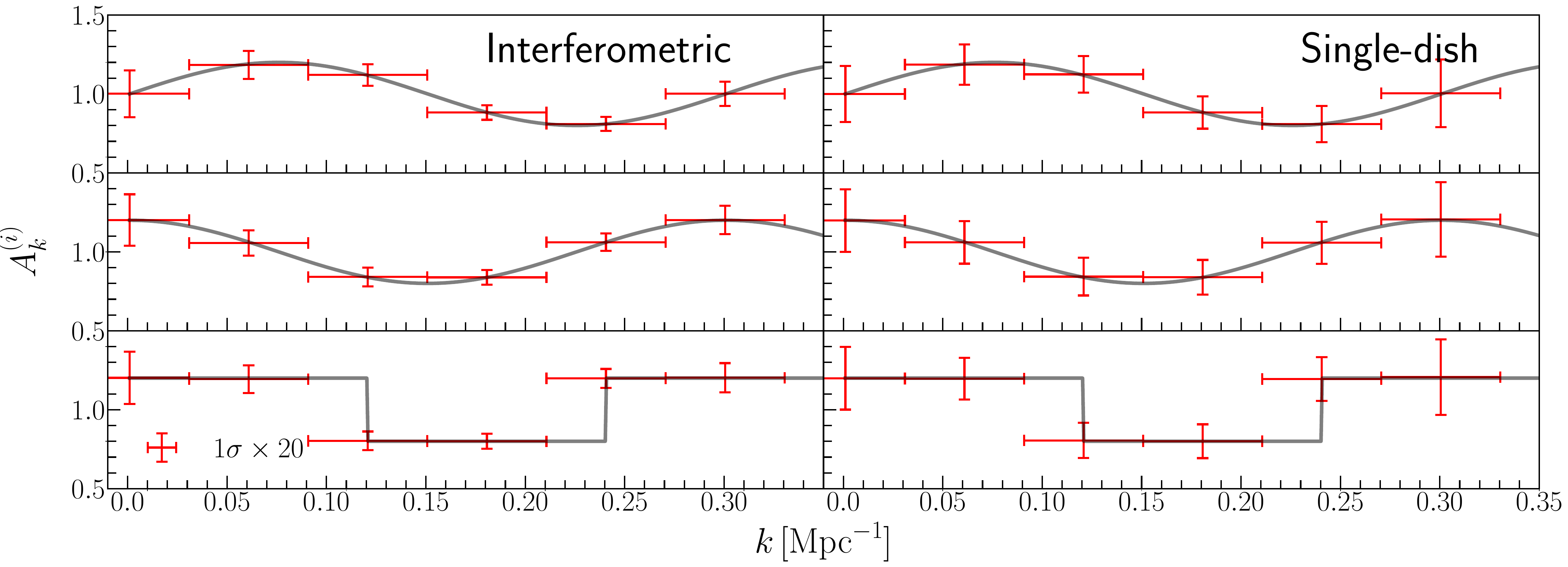}
\caption{Reconstruction validations of different HI signals. HI signals with shapes and amplitudes modulated in different ways can be correctly reconstructed from the Bayesian analysis in Eq. (\ref{MLE}) for interferometric IM experiments (left) and for single-dish-like IM experiments (right), respectively. The fiducial value of $A_{k}^{(i)}=1$ without modulation. }\label{kpdfs}
\end{figure*}

\textit{Bayesian analysis framework.---}The simulated sky signal can be expressed as ${\bf C}^{\rm obs}={\bf C}^{\rm fg}+{\bf C}^{\rm HI}+{\bf C}^{\rm BAO}+{\bf N}$, in which the foreground is strongly correlated among different frequency bands whereas the HI is only correlated at neighboring bands. Here ${\bf C}$ denotes a power spectrum vector. The power-spectrum space thus can be split into a cross-power ($\times$) spectrum subspace and the auto-power ($\circ$) spectrum subspace by imposing a frequency-correlation length $\Delta\nu<\Delta\nu_{\ast}$ [MHz]. The frequency coherence length considered in this work is $\Delta\nu_{\ast}$ = 20 [MHz]. We adopt a principal component analysis (PCA) scheme to remove the strongly correlated components that largely correspond to foreground contamination. The noise-debiased band powers are decomposed as ${\bf  C}^{{\rm obs}}-{\bf N}={\bf  V}{\bf \Lambda} {\bf  V}^T$ where ${\bf  V}$ and ${\bf \Lambda}$ are the corresponding eigenvectors and eigenvalues. By forming a rank-$m$ subspace which contains the largest $m$ eigenvalues ${\bf\Lambda}^{\prime}$ and the eigenvectors ${\bf V}^{\prime}$, we can directly remove the radio foreground ${\bf C}^{\rm fg,\ast}={\bf V}^{\prime}{\bf\Lambda}^{\prime} {\bf V}^{\prime,T}$ in the power-spectrum space as ${\bf  C}^{\rm clean, \ast}={\bf  C}^{{\rm obs}}-{\bf C^{\rm fg,\ast}}={\bf  C}^{{\rm obs}}-{\bf V}^{\prime}{\bf\Lambda}^{\prime} {\bf V}^{\prime,T}$, achieving a $\mathcal{O}(10^{-6})$ level removal of the overall foreground contamination. We have tested this procedure with different $m$ values and find that a minimum value $m=2$ can achieve a satisfying foreground residual level as demonstrated in Fig. \ref{fgfit}.

The HI power is almost concentrated in the auto-power subspace as shown in Fig. \ref{fgfit}, thereby we can perform the Bayesian analysis as
\begin{eqnarray}
-2\ln\mathcal{L}[{\bf P}|{\bf C^{(\rm obs)}}]&=& [{\bf d}^{\circ}]^T[\Delta\mathcal{C}^{(\rm fg,\circ)}+\mathcal{C}^{(\rm HI,\circ)}+\mathcal{N}^{(\circ)}]^{-1}[{\bf  d}^{\circ}]\nonumber\\
&+&{\ln} |\Delta\mathcal{C}^{(\rm fg,\circ)}+\mathcal{C}^{(\rm HI,\circ)}+\mathcal{N}^{(\circ)}|,\nonumber\\
\label{MLE}
\end{eqnarray}where we define the differenced data vector in the auto-power subspace ${\bf d}^{\circ}={\bf C^{(\rm obs,\circ)}}-{\bf C^{(\rm fg, \ast, \circ)}}-{\bf C^{(\rm HI, \circ)}}-{\bf C^{(\rm sys, \circ)}}-{\bf N}$ and $|...|$ denotes the determinant. The covariance matrix is defined as $\mathcal{C}_b\delta_{bb^{\prime}}=\langle({\bf C}_b-{\bf \bar C}_b)({\bf C}_{b^{\prime}}-{\bf \bar C}_{b^{\prime}})^T\rangle$ which can be calculated by the Knox formula [see Eq. (\ref{knox}) in \textit{Supplemental Material}]. The foreground residual is quantified by a parameter $\eta$ which is set to $10^{-6}$ as a result of the power-spectrum PCA removal of the foreground contaminants, thus the corresponding covariance is $\Delta\mathcal{C}^{(\rm fg,\circ)}=\eta^2\mathcal{C}^{(\rm fg,\ast,\circ)}$. $\mathcal{C}^{(\rm HI)}$ and $\mathcal{N}$ are covariance matrices for the HI and instrumental noise. The parameter set of the radio sky model is $\textbf{P}=\{{\textbf A}_k, A_{\rm osc}, \alpha, {\bf \epsilon}\}$. There are $N_k$ parameters for modeling the neutral hydrogen power spectra, two parameters for the BAO power spectra, and three parameters for the systematic effects. So the total parameter number is $N_k+5$. In this work, we use the latest Planck cosmological parameters~\cite{2020A&A...641A...6P}.

This PSDFT method differs from the maximum likelihood spectral matching developed for the CMB data analyses~\cite{2003MNRAS.346.1089D} in that foreground contaminants that are a few orders of magnitude stronger at radio bands have to be first subtracted before the Bayesian analysis can be applied to the HI signal reconstruction. Moreover, this PSDFT method can analyze orders of magnitude more frequency bands than the CMB by taking advantage of the distinct frequency-correlation patterns of the HI signals.

The derivative of $\ln{|\mathcal{C}|}$ with respect to the parameters is proportional to ${\rm Tr}({\mathcal{C}^{-1}\partial \mathcal{C}/\partial {\bf P}})=\langle u\mathcal{C}^{-1}\partial \mathcal{C}/\partial {\bf P}\mathcal{C}^{-1}u\rangle$ with $u$ drawn from the covariance $\mathcal{C}$. This term contains information about the mean of the ensemble so is insensitive to the component reconstructions and is neglected. The instrumental noise covariance is considered as a diagonal matrix $\mathcal{N}$ for a constant noise power spectrum $N_{\ell}$. We note that systematic issues such as cross-talk of detectors and $1/f$ noise can generate non-negligible off-diagonal elements in the noise covariance and defer the discussions of these realistic effects to future work. 

\textit{Results and validations.---}In this work we simulate mock data within $925 <\nu<1075$ MHz with $N_{\nu}=30$ frequency bands and bin the power spectra within $50<\ell<500$ with $N_b=6$ band powers for an IM survey with $f_{\rm sky}=0.35$. To model the HI power spectra, we select $N_k=6$ $k$-modes among $0.001<k<0.6$ ${\rm Mpc}^{-1}$ which correspond to the multipole $\ell$ range $50<\ell<500$ [see Fig. \ref{kwavelets} in \textit{Supplemental Material}]. For the BAO model in Eq. (\ref{baomodel}), the pivot scale is $k_0=0.1{\rm Mpc}^{-1}$, the sound horizon scale $k_A=0.1{\rm Mpc}^{-1}$, the spectral index $n_d=1.5$ and the overall amplitude $A_{\rm osc}=1$. Also, we assume a reference frequency $\nu_{\rm ref}=1000$ MHz and foreground removal efficiency $\eta=10^{-6}$. We consider an interferometric IM experiment with a low noise level $N_{\ell}=10^{-10}{\rm \mu K}\mbox{-}{\rm arcmin}$ and a high resolution $\theta_{\rm ref}=5^{\prime}$ and a single-dish-like IM experiment with a noise level $N_{\ell}=10^{-8}{\rm \mu K}\mbox{-}{\rm arcmin}$ and low resolution $\theta_{\rm ref}=30^{\prime}$. 

We sample the parameter space $\textbf{P}=\{{\textbf A}_k, A_{\rm osc}, \alpha, {\bf \epsilon}\}$ using the Bayesian probability in Eq. (\ref{MLE}). All the parameters of the neutral hydrogen and BAO models are correctly reconstructed with high significance, implying that the faint HI signals can be precisely extracted from the raw data that are highly contaminated by foregrounds (Figs. \ref{pdf_ref} and \ref{pdf_dish} in \textit{Supplemental Material}). The much fainter BAO signals at different angular scales can also be recovered simultaneously with high signal-to-noise ratios. The PSDFT approach enables us to simultaneously decompose the raw data into individual components as shown in Fig. \ref{kpdfs2}. Especially, the HI signals without BAO wiggles can be obtained from the raw data. This can be regarded as the first ``de-BAO'' analysis which would benefit future analyses requiring no BAO contributions. 

We have tested different HI signals as shown in Fig. \ref{kpdfs} for both interferometric (left) and single-dish-like (right) IM experiments, respectively. It is seen that all components are correctly reconstructed, indicating that the PSDFT approach is independent of the HI models which can vary with different shapes and amplitudes. The residuals of the foreground contamination and reconstructed HI angular power spectra are found to be negligible in Fig. \ref{fgfit}. 

With rapid developments of interferometric detection techniques, future intensity mapping surveys with neutral hydrogen, carbon monoxide, and the 158$\mu{\rm m}$ fine structure line of single ionized carbon~\cite{2005ApJ...625..575S}, will perform multifrequency or tomographic observations. This proof-of-concept study demonstrates that the PSDFT approach can be applied to future multifrequency experiments and opens up a window for processing large volumes of intensity mapping data in the power spectrum domain. There are certain realistic complexities such as the spatial non-Gaussian structures, non-Gaussian beam profiles~\cite{2021MNRAS.506.5075M} and foreground polarization leakage~\cite{2004ApJS..152..129V, 2016ApJ...833..289L, 2017ApJ...848...47N}. All of these in principle can be directly extended within this PSDFT framework.

\acknowledgments
\textit{Acknowledgments.---}
We are grateful for the helpful discussions with Gilbert Holder. This work is supported by the National Key R\&D Program of China Grant No. 2022YFC2204603 and by the starting grant of USTC. We acknowledge the use of the \healpix~\cite{2005ApJ...622..759G} and \emcee~\cite{2013PASP..125..306F} packages.

\bibliography{refs}
\newpage

\medskip

\onecolumngrid

\setcounter{page}{1}
\renewcommand{\thepage}{S\arabic{page}} 
\renewcommand{\thetable}{S\Roman{table}}  
\renewcommand{\thefigure}{S\arabic{figure}}
\renewcommand{\theequation}{S\arabic{equation}}
\setcounter{section}{1}
\setcounter{figure}{0}
\setcounter{table}{0}
\setcounter{page}{1}
\setcounter{equation}{0}
\newpage

\begin{center}
\textbf{\large \psdfttitle} \\
\vspace{0.05in}
{ \it \large Supplemental Material}\\
\vspace{0.05in}
{ Chang Feng, Filipe B. Abdalla}
\end{center}

Brightness temperature fluctuations of neutral hydrogen are $\delta T_b=T_0\delta_b=T_0b_{\rm HI}\delta_c$, where $b_{\rm HI}$ is the bias and the averaged brightness temperature~\cite{2015ApJ...803...21B} is $T_0=0.566h/E(z)(\Omega_{\rm HI}/0.003)(1+z)^2{\rm mK}$ where $E(z)=H(z)/H_0$ is the normalized Hubble constant, and the Hubble constant today $H_0$ is defined as $H_0=100h\,{\rm km\,s}^{-1}\,{\rm Mpc}^{-1}$. $\delta_b$ and $\delta_c$ are density contrast of baryon and dark matter, respectively. In this work, we ignore the redshift distortions and assume the galaxy bias and HI fraction are constant, i.e., $b_{\rm HI}=1$ and $\Omega_{\rm HI}=6.2\times10^{-4}$~\cite{2013MNRAS.434L..46S} which is consistent with the latest measurement $8.6\times 10^{-4}$ from the HI and galaxy cross-power spectrum~\cite{2023MNRAS.518.6262C}. Therefore, the HI power spectrum is $P_{\rm HI}(k)=T^2_0(z)b^2_{\rm HI}P(k)$ where $P(k)$ is a three-dimensional matter power spectrum. In this work, we generate the matter power spectra $P(k)$ with and without BAO wiggles by computing transfer functions $F^{w}(k)$ and $F^{\rm nw}(k)$, respectively~\cite{1998ApJ...496..605E}. Specifically, the matter power spectrum without BAO wiggles is 
\begin{equation}
P^{\rm nw}(k)\sim k^{n_s}F^{2,({\rm nw})}(k)G^2(z)
\end{equation}
which is normalized by $\sigma_8$ today~\cite{1998ApJ...496..605E, 2009ApJS..180..330K}. Here $n_s$ is the spectral index of the curvature perturbation and $G(z)$ is the cosmological structure growth factor calculated assuming a fiducial cosmological model determined by the latest Planck cosmological parameters~\cite{2020A&A...641A...6P}.

An IM experiment measures the radio signals at a finite frequency bandwidth $\Delta\chi_{\nu}=(\Delta\nu/{0.1\, {\rm MHz}})\sqrt{(1+z_{\nu})/10}(\Omega_m h^2/0.15)^{-1/2}\,{\rm Mpc}$~\cite{2013ApJ...779..124M} which corresponds to a narrow redshift coverage
\begin{equation}
W_{\nu}(z)=\frac{1}{\sqrt{2\pi\Delta\chi^2_{\nu}}}e^{-\frac{1}{2}\Big (\frac{\chi-\chi_{\nu}}{\Delta\chi_{\nu}}\Big)^2},
\end{equation}where $z_{\nu}=\nu_{\rm 21cm}/\nu-1$ and $\nu_{\rm 21cm}=1420 {\rm Hz}$. 

A narrow-band observation is an integral of the brightness temperature fluctuations within a finite redshift range. In this work, we adopt a $k$-wavelet decomposition of the angular power spectrum at frequencies $\nu_A$ and $\nu_B$, i.e., 
\begin{eqnarray}
&&\int \frac{d\chi}{\chi^2}W_{\nu_A}(z)W_{\nu_B}(z)[T_0(z)b_{\rm HI}]^2P^{\rm nw}(k)=\displaystyle\sum_{i=1}^{N_k}A^{(i)}_kC^{(i),\nu_A\times\nu_B}_{\ell},\nonumber\\
\end{eqnarray}
where the $k$ mode is
\begin{eqnarray}
C^{(i),\nu_A\times\nu_B}_{\ell}&=&\int \frac{d\chi}{\chi^2}W_{\nu_A}(z)W_{\nu_B}(z)[T_0(z)b_{\rm HI}]^2\eta_i(k)P^{\rm nw}(k).
\end{eqnarray} and $\eta_i(k)$ is a top-hat window at the $i$-th $k$-mode satisfying the unity relation $\sum_{i=1}^{{N_k}}\eta_i(k)=1$. We show a representative plot for different $k$ modes of the HI power spectrum at $\nu$ = 980 MHz in Fig. \ref{kwavelets} (left). For an IM experiment with $N_b$ angular broad bands within $\ell_{\rm min}<\ell<\ell_{\rm max}$, we use the Limber approximation $k=\ell/\chi$ to map the detailed $\ell\mbox{-}k$ space such as Fig. \ref{kwavelets} (right) so specific $k$ wavelets can be determined.

\begin{figure*}
\includegraphics[width=8cm, height=6cm]{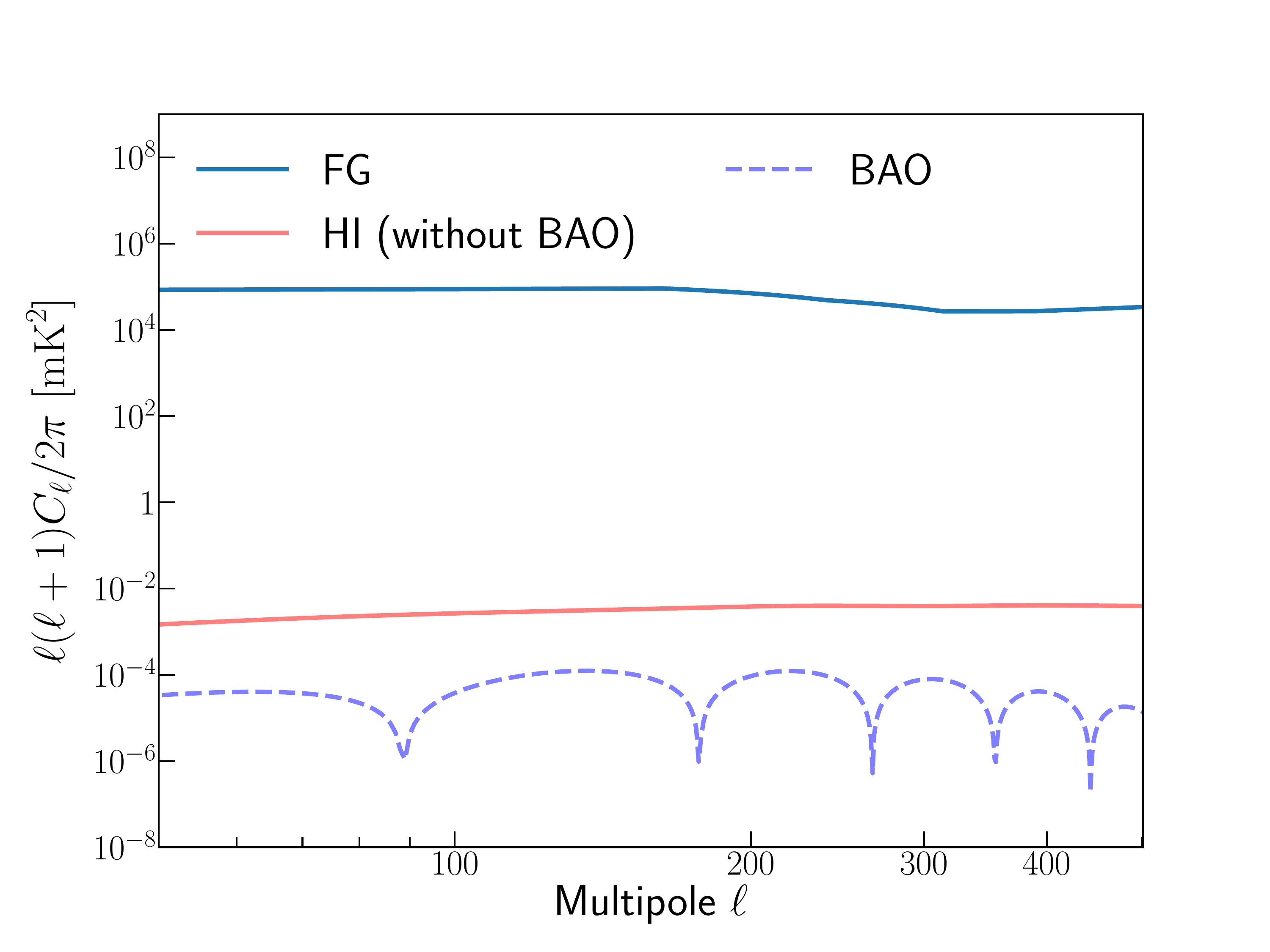}
\includegraphics[width=8cm, height=6cm]{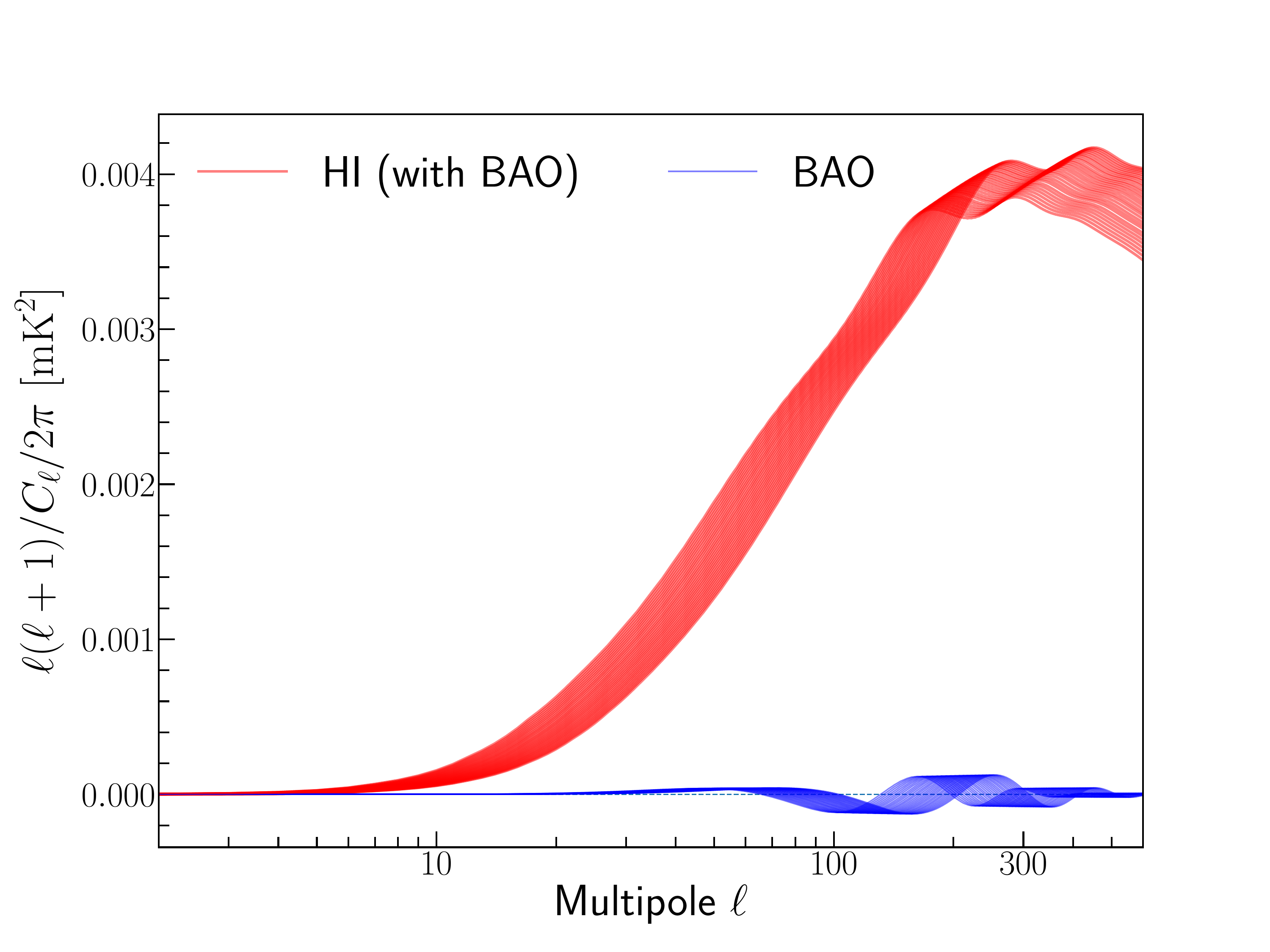}
\caption{(left) Component power spectra at $\nu=$ 980 MHz. The foreground contaminants including synchrotron emission, free-free emission, anomalous microwave emission, and the cosmic microwave background are a few orders of magnitude brighter than the HI signals, and the HI signals are about two orders of magnitude stronger than the BAO. The absolute values of the BAO component are shown in the dashed line. (right) All HI power spectra at frequencies between 925 and 1075 MHz. The BAO components are included in the HI signals. In this work, we ignore the redshift distortions and assume the galaxy bias and HI fraction are constant, i.e., $b_{\rm HI}=1$ and $\Omega_{\rm HI}=6.2\times10^{-4}$~\cite{2013MNRAS.434L..46S}. For BAO modeling, we adopt the phenomenological model in Eq. (\ref{baomodel}). }\label{components}
\end{figure*}

A shift parameter $\alpha$ can be incorporated into the $f(k)$ function in Eq. (\ref{baomodel}) as $f(k^{\prime})=f(k/\alpha)$ and is assumed to be a small deviation from unity. Therefore, we expand $f(k^{\prime})$ around $\alpha_0$ to the $N_{\alpha}$-th order as~\cite{2023MNRAS.519..799H}
\begin{equation}
f(k;\alpha)=f(k;\alpha_0)+\displaystyle\sum_{n=1}^{N_{\alpha}}\frac{1}{n!}\frac{d^{(n)}f}{d\alpha^{(n)}}{\Big |}_{\alpha=\alpha_0}(\alpha-\alpha_0)^n.
\end{equation}
In this work, we choose $\alpha_0=1$ and find that $N_{\alpha}=12$ can yield good convergence. Following the neutral hydrogen model in the main text, the matter power spectrum with BAO is $P^{\rm w}(k)=[1+f(k)]P^{\rm nw}(k)$ where the second term is the BAO-induced contribution. We only apply the shift parameter to the BAO-induced features $f(k)$ and keep the smooth power spectrum $P^{\rm nw}(k)$ unchanged~\cite{2021MNRAS.506.2638K}. Thereby, the full parametrization of the HI power spectrum at two frequencies $\nu_A$ and $\nu_B$ is
\begin{eqnarray}
C^{\nu_A\times\nu_B}_{b}&=&\displaystyle\sum_{i=1}^{N_k}A^{(i)}_kC^{(i),\nu_A\times\nu_B}_{b}+A_{\rm osc}\Big[C^{\nu_A\times\nu_B,{\rm BAO},(0)}_{b}+\displaystyle\sum_{n=1}^{N_{\alpha}}\frac{1}{n!}(\alpha-\alpha_0)^n\Delta^{\nu_A\times\nu_B,{\rm BAO},(n)}_b\Big],\nonumber\\
\end{eqnarray}
where the $n$-th BAO perturbation term is
\begin{eqnarray}
\Delta^{\nu_A\times\nu_B,{\rm BAO},(n)}_b&=&\int \frac{d\chi}{\chi^2}W_{\nu_A}(z)W_{\nu_B}(z)[T_0(z)b_{\rm HI}]^2\frac{d^{(n)}f}{d\alpha^{n}}{\Big |}_{\alpha=\alpha_0}P^{\rm nw}(k),\label{baopert}\end{eqnarray}
and the overall amplitude is $A_{\rm osc}$. As seen from Figs. \ref{components}, the BAO-induced power spectrum is about two orders of magnitude smaller than the HI.

With all the components mentioned above, the angular band powers of the radio sky can be described as
\begin{eqnarray}
&&{\bf C}^{\nu_A\times\nu_B}_{b}({\textbf A}_k, A_{\rm osc}, \alpha, {\bf \epsilon})\nonumber\\&=&{\bf C}^{\rm fg}+{\bf C}^{\rm HI}+{\bf C}^{\rm BAO}+{\bf C}^{\rm sys}+{\bf N}\nonumber\\
&=&\displaystyle\sum_{i}^{N_f}C_{b}^{{\rm fg}(i), \nu_A\times\nu_B}+\displaystyle\sum_{i=0}^{N_k}A^{(i)}_kC^{(i),\nu_A\times\nu_B}_{b}+A_{\rm osc}\Big[C^{\nu_A\times\nu_B,{\rm BAO},(0)}_{b}+\displaystyle\sum_{n=1}^{N_{\alpha}}\frac{1}{n!}(\alpha-\alpha_0)^n\Delta^{\nu_A\times\nu_B,{\rm BAO},(n)}_b\Big]\nonumber\\
&+&\displaystyle\sum_{i=-1}^{1} C^{\nu_A\times\nu_B, {\rm bb},(i)}_b+N^{\nu_A\times\nu_B}_{b}.\label{sky}
\end{eqnarray}

\begin{figure}
\includegraphics[width=8cm, height=5cm]{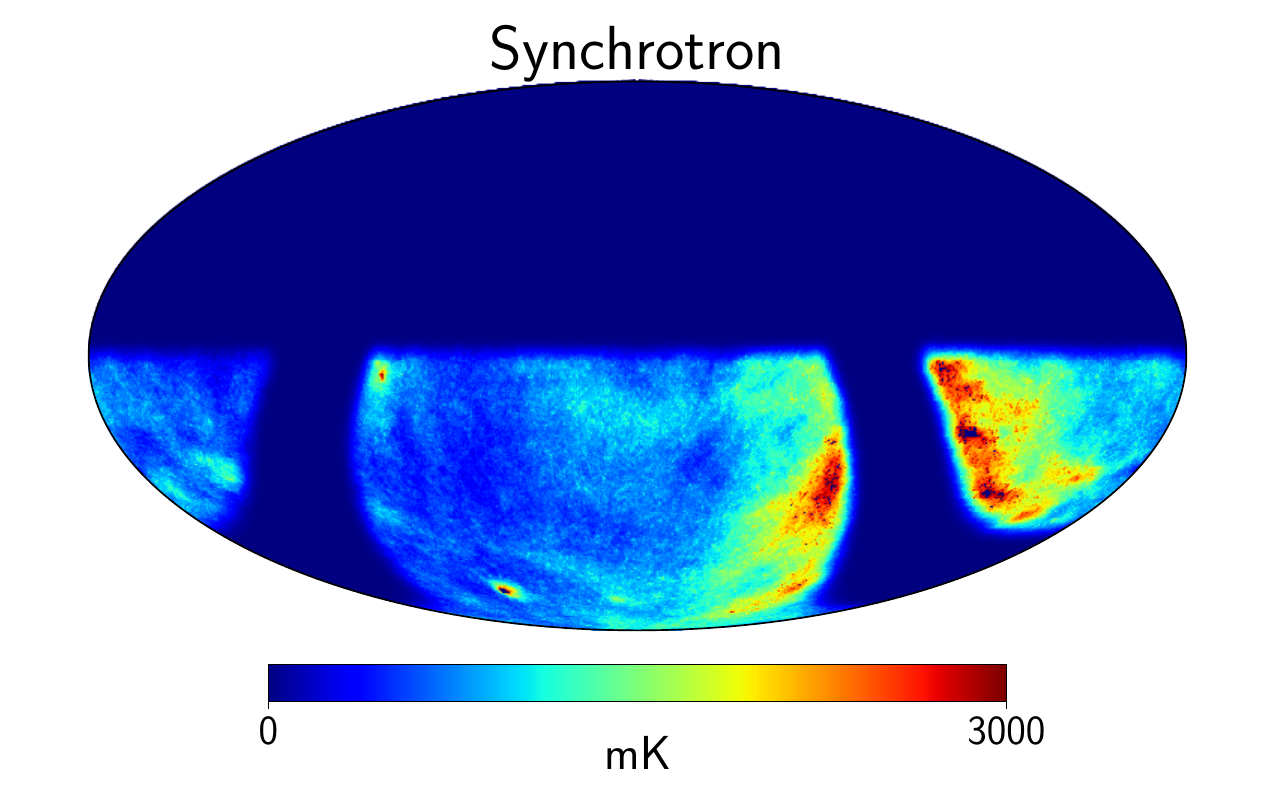}
\includegraphics[width=8cm, height=5cm]{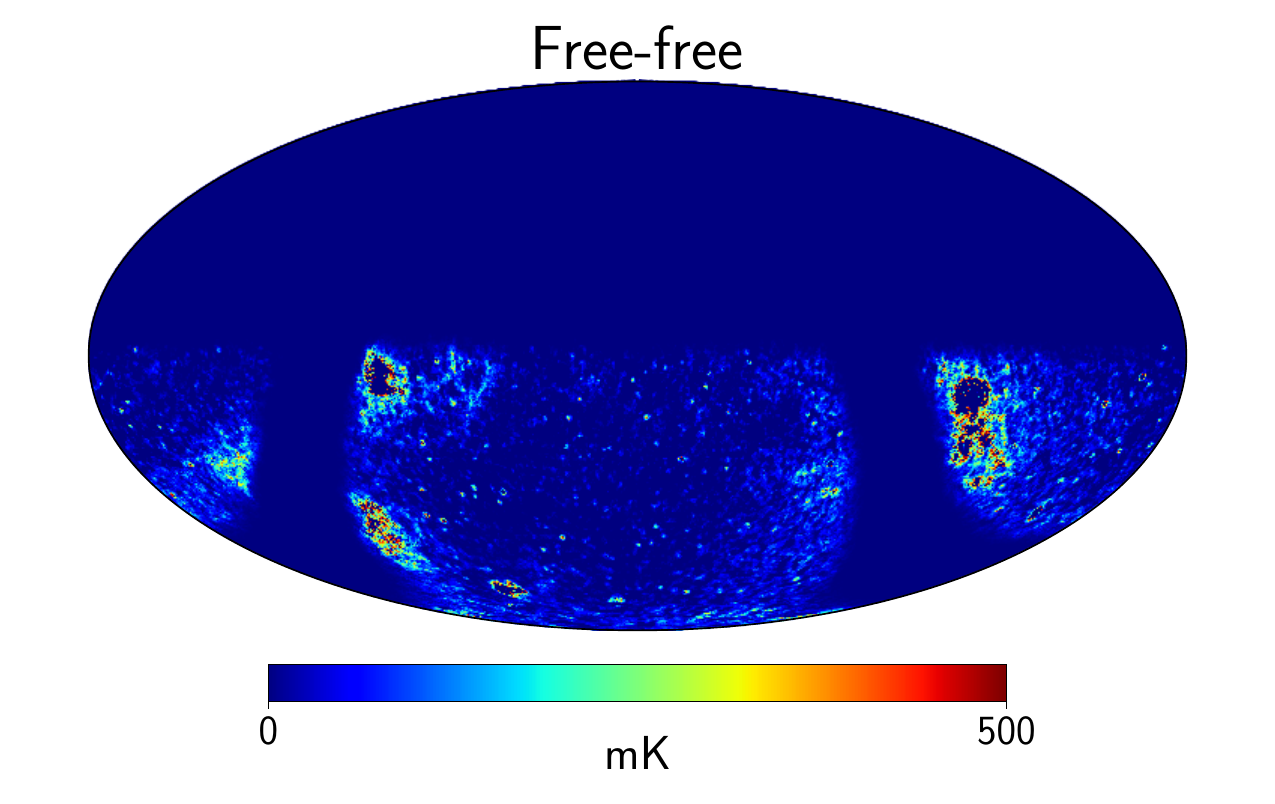}
\includegraphics[width=8cm, height=5cm]{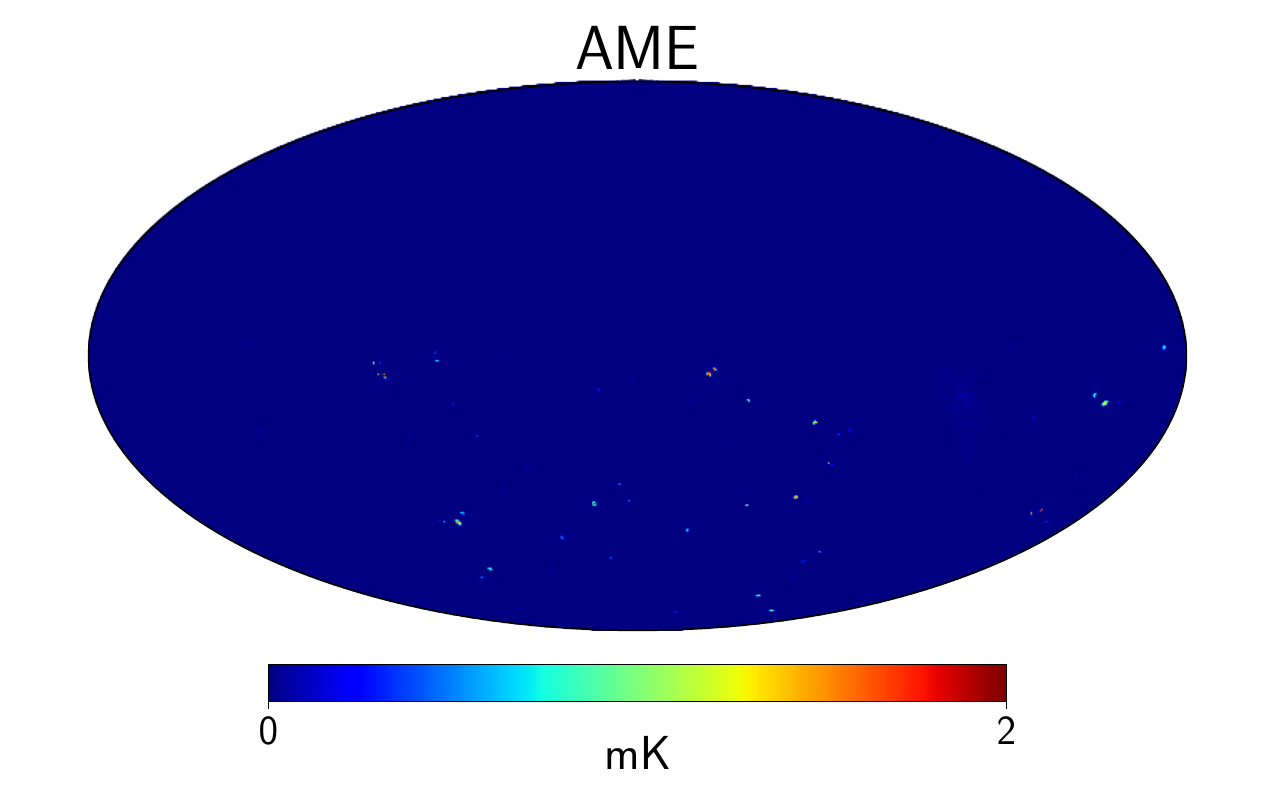}
\includegraphics[width=8cm, height=5cm]{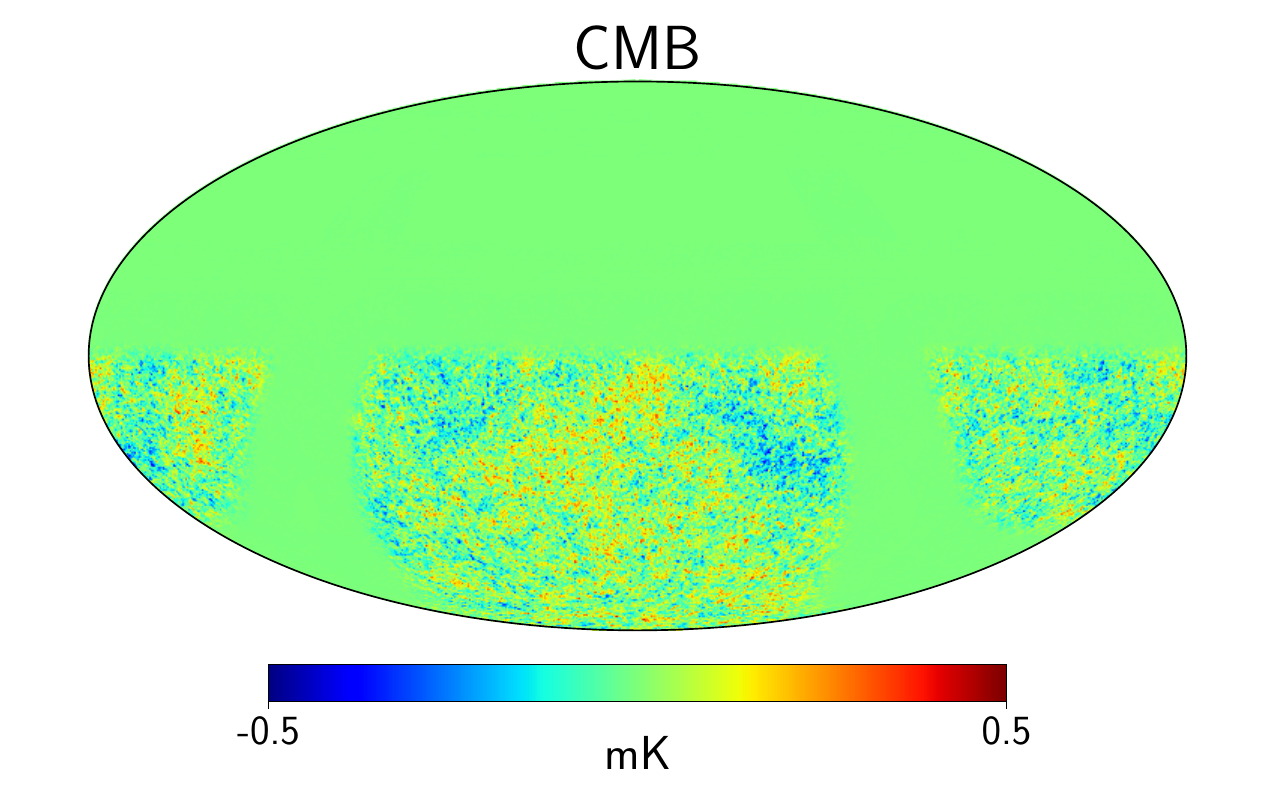}
\includegraphics[width=8cm, height=5cm]{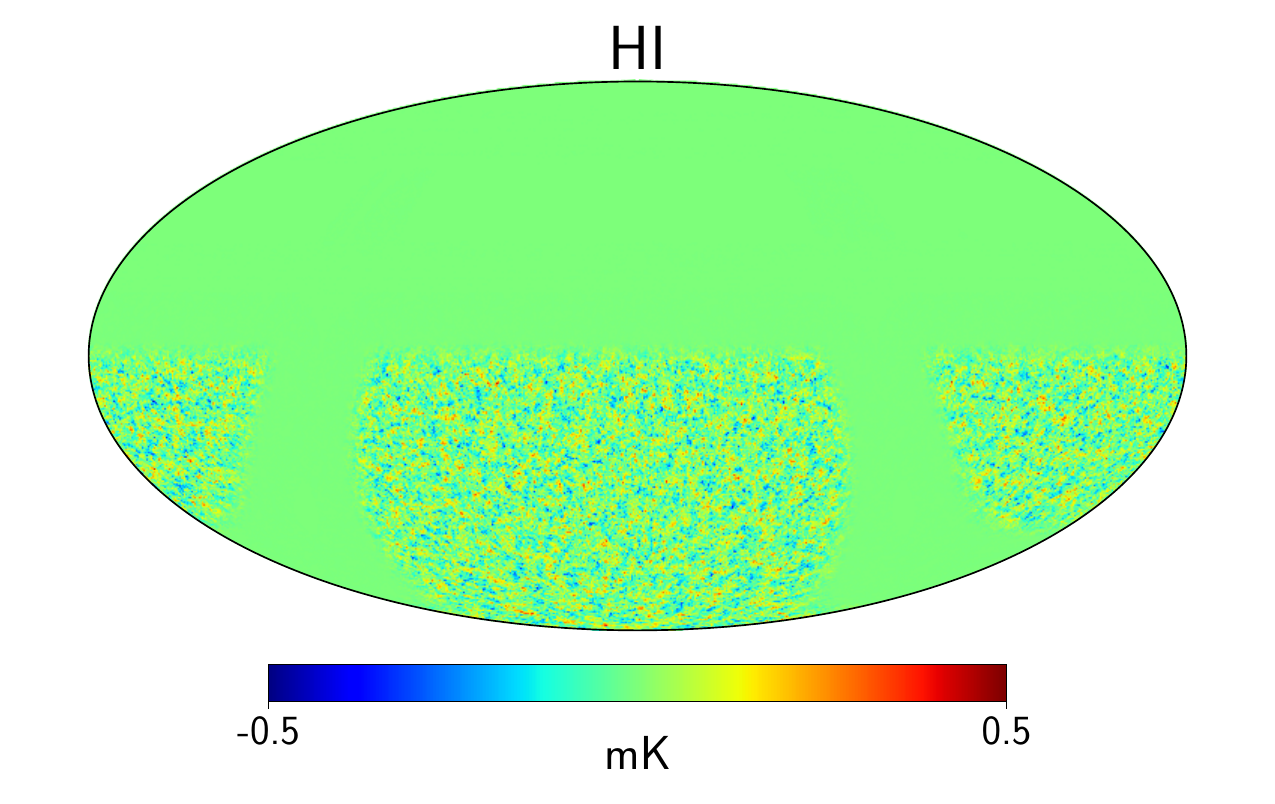}
\includegraphics[width=8cm, height=5cm]{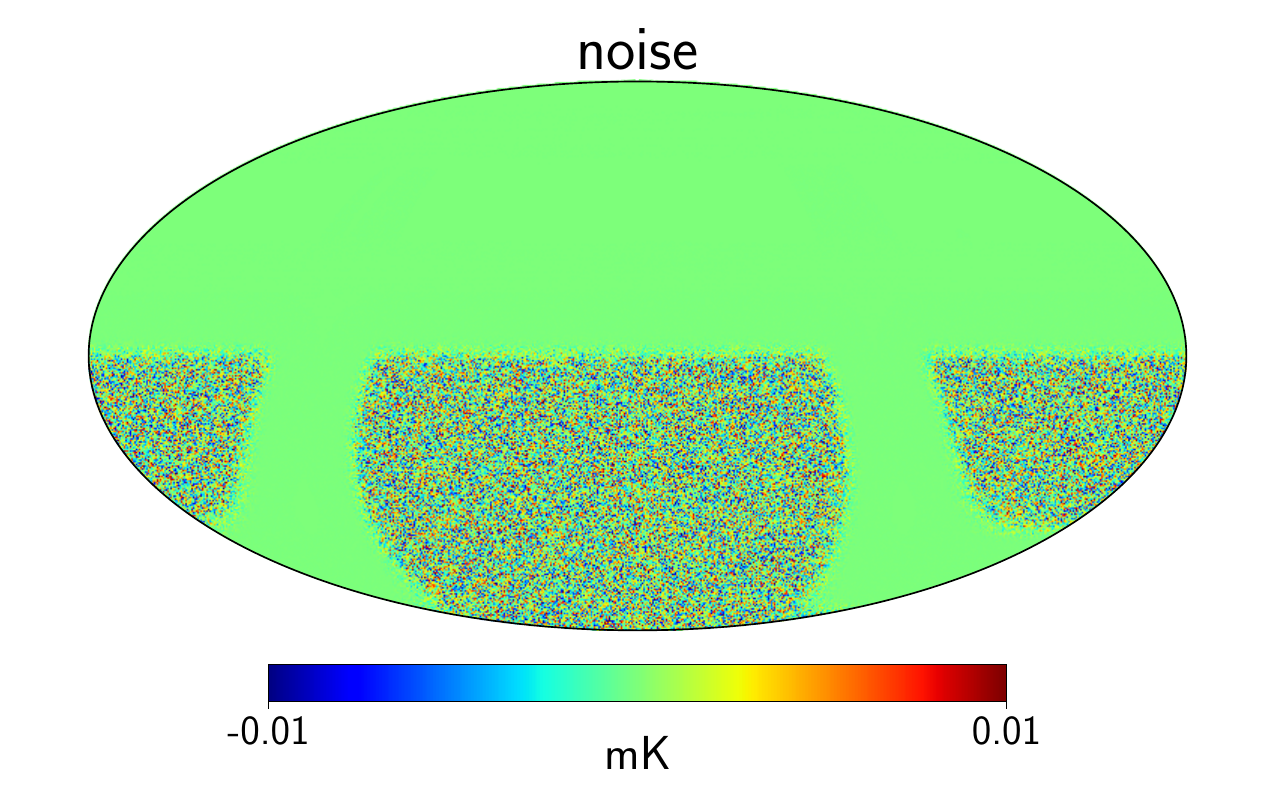}
\caption{Simulated foreground contaminants, HI signals, and noise components at frequency $\nu$ = 1000 MHz in equatorial coordinates. Foreground contaminants are generated by the Python Sky Model with a \healpix pixelization at a resolution $N_{\rm side}$ = 512. The mock data assumes a southern hemisphere observation with the Galactic plane excluded, covering 35\% of the full sky. The noise realization shown in this plot corresponds to the interferometric IM survey. ``AME'' and ``CMB'' refer to anomalous microwave emission and the cosmic microwave background.}\label{mock}
\end{figure}

The power spectra of the summed foreground contaminants from $N_f$ components and HI fluctuations are shown in Fig. \ref{components} (left). The corresponding map realizations generated by the Python Sky Model~\cite{2017MNRAS.469.2821T} are shown in Fig. \ref{mock}.

\begin{figure*}
\includegraphics[width=8cm, height=6cm]{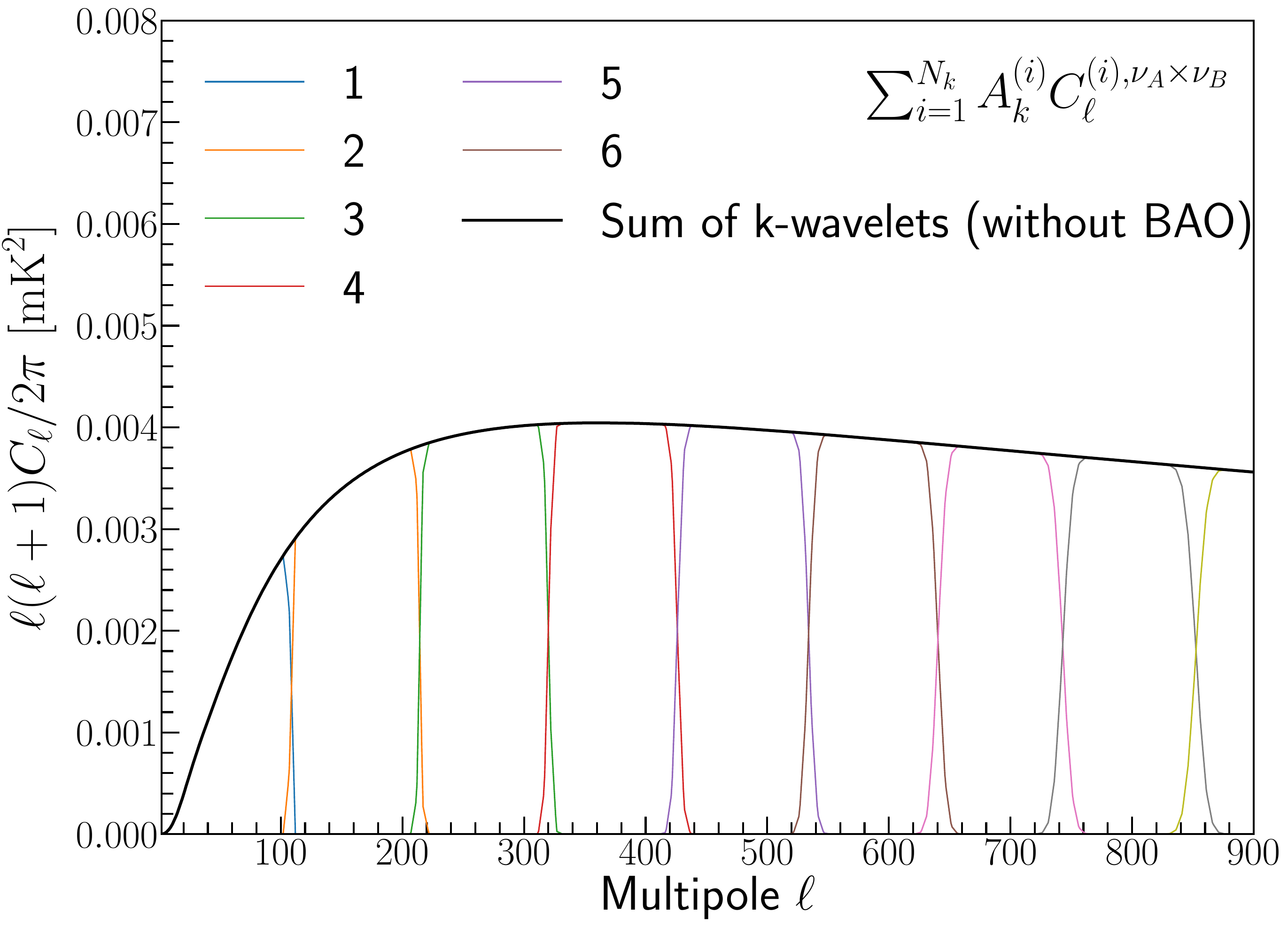}
\includegraphics[width=8cm, height=6cm]{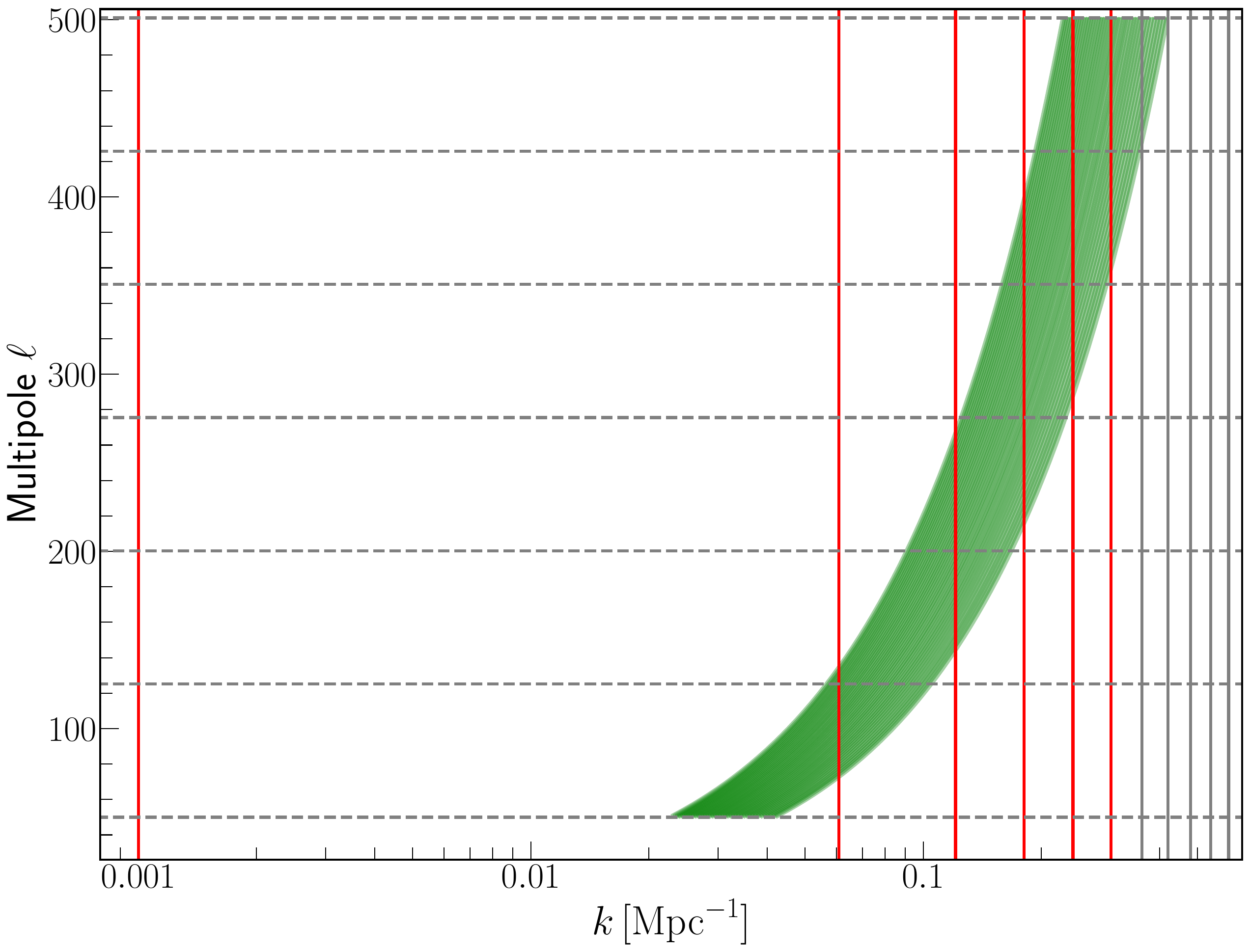}
\caption{(left) Representative $k$-mode HI power spectra at $\nu=$ 980 MHz. We only use the power spectra at $50<\ell<500$ which correspond to $k$-wavelets 1-6 as indicated by the $\ell\mbox{-}k$ relation. (right) The $\ell\mbox{-}k$ space mapping for the power spectra at $50<\ell<500$ for all frequencies. The red vertical lines indicate the $k$-wavelets that are taken into account in this work.}\label{kwavelets}
\end{figure*}

\begin{figure*}
\includegraphics[width=16cm, height=7cm]{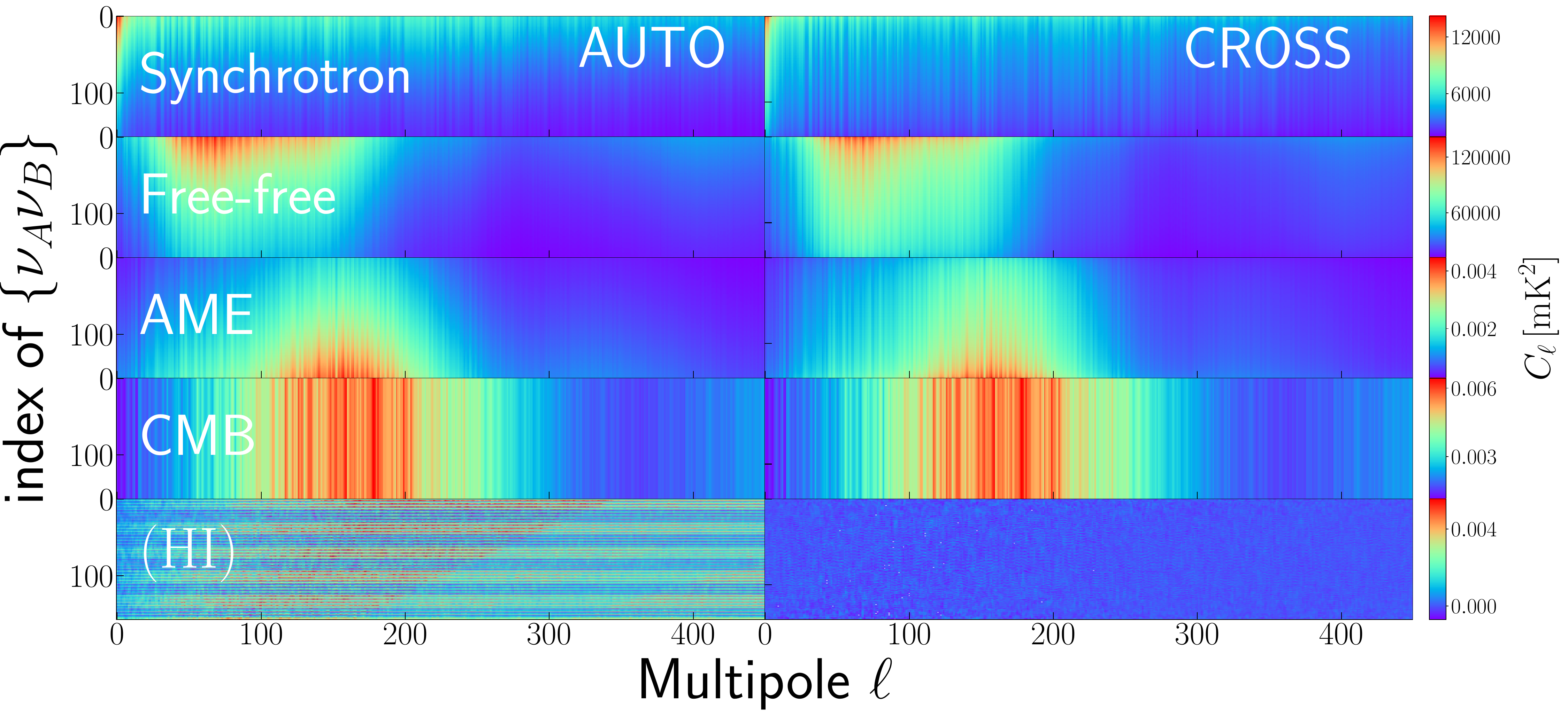}
\caption{Raw power spectra of different components at all frequencies with respect to different angular scales. The power-spectrum space can be split into a cross-power ($\times$) spectrum subspace labeled as ``CROSS'' and the auto-power ($\circ$) spectrum subspace labeled as ``AUTO'' by imposing a frequency-correlation length $\Delta\nu<\Delta\nu_{\ast}$ [MHz]. The frequency coherence length considered in this work is $\Delta\nu_{\ast}$ = 20 [MHz]. }\label{psdftRAW}
\end{figure*}

\begin{figure*}
\includegraphics[width=15cm, height=10cm]{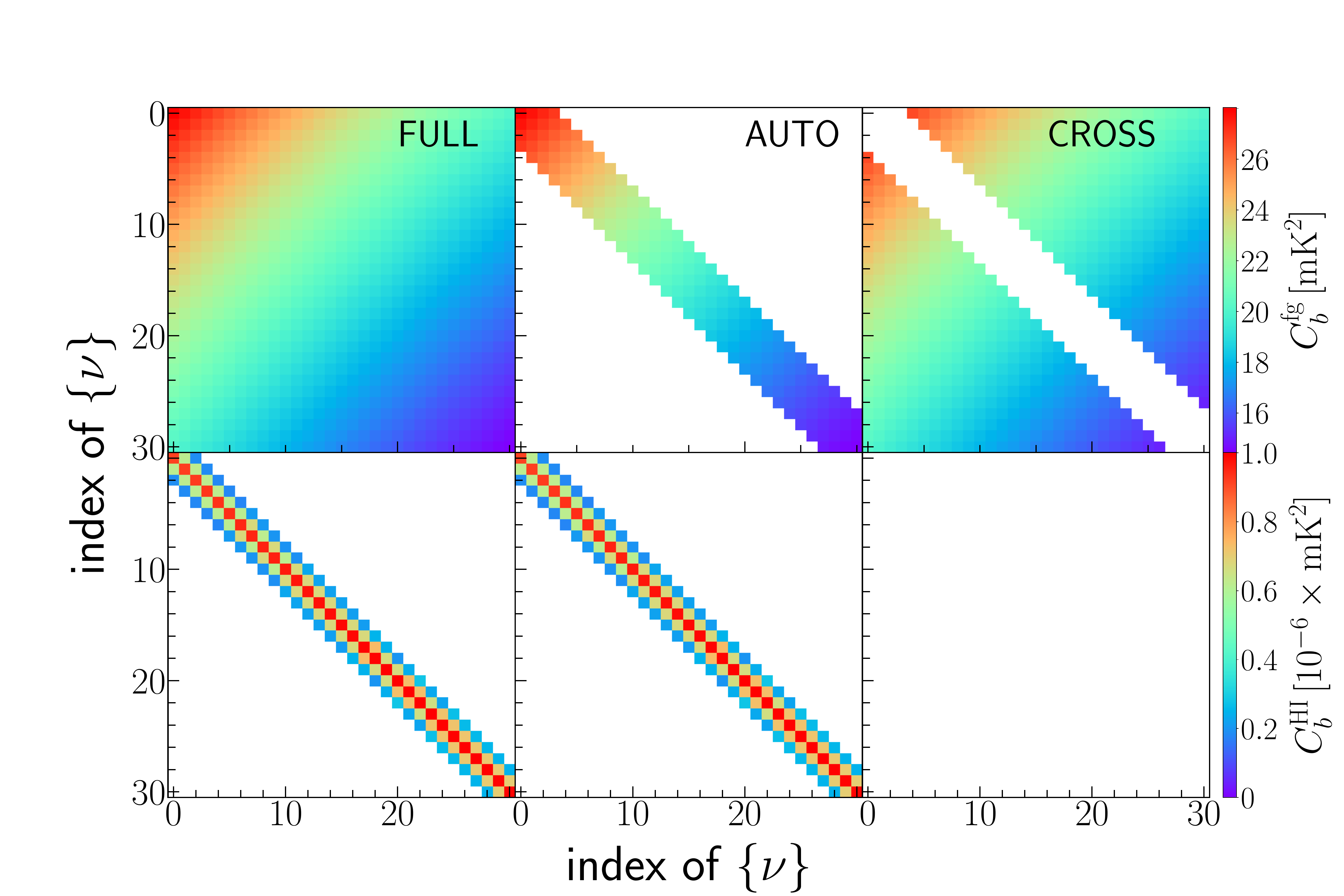}
\caption{Angular band powers at all frequencies for a single broadband at $b$ = 162. The axis is arranged by the indices of all the frequencies $\{\nu\}$. The description of ``CROSS'' and ``AUTO'' is the same as Fig. \ref{psdftRAW}. The label ``FULL'' denotes the power-spectrum space without decomposition. The first row shows the power-spectrum space and its decomposition for the overall foreground contamination. The second row corresponds to the HI power spectra. As seen from the last subplot, the HI signal is completely null at the cross-power spectrum subspace.}\label{psdft}
\end{figure*}

\begin{figure*}
\includegraphics[width=8cm, height=6.7cm]{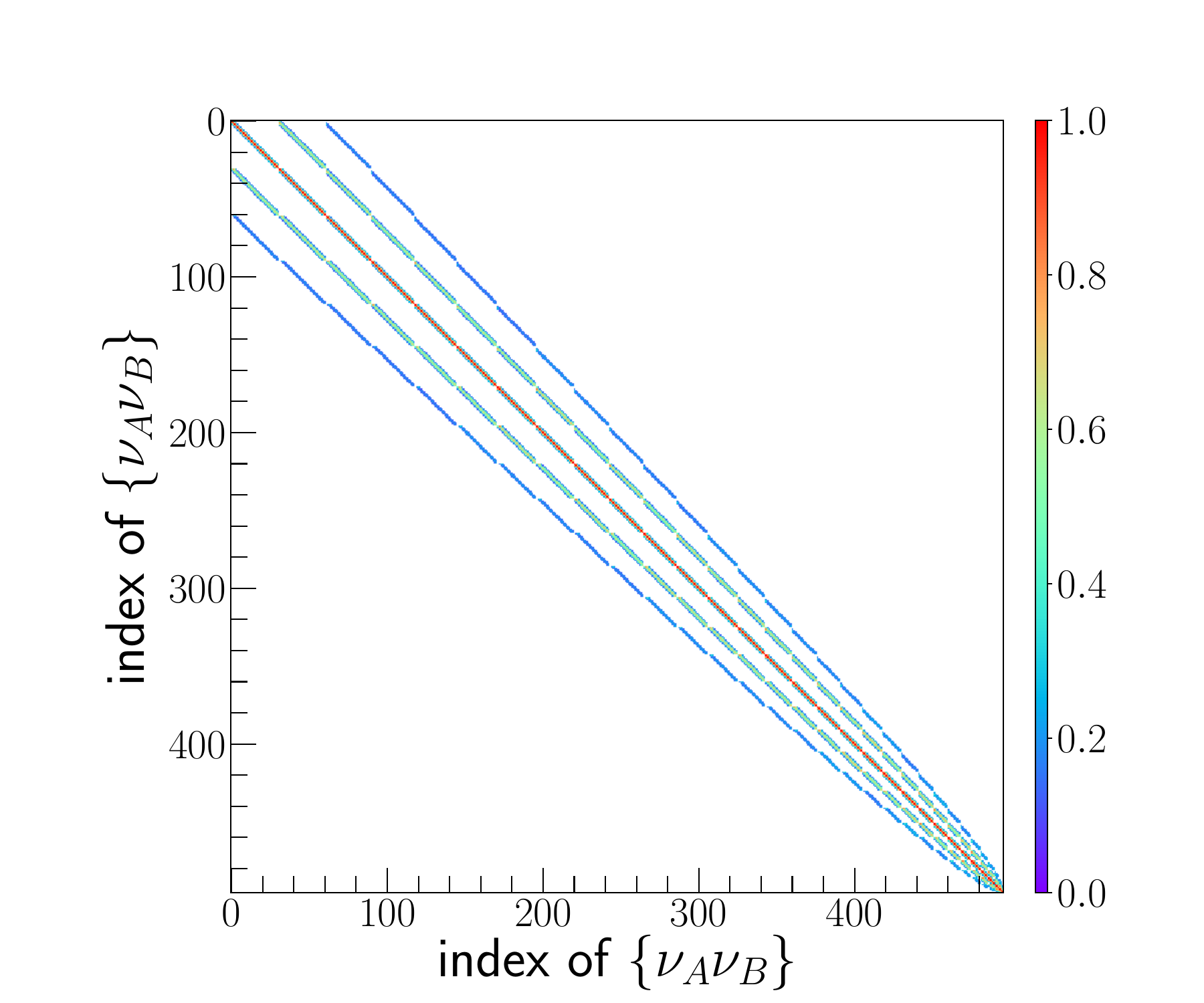}
\includegraphics[width=8cm, height=6.7cm]{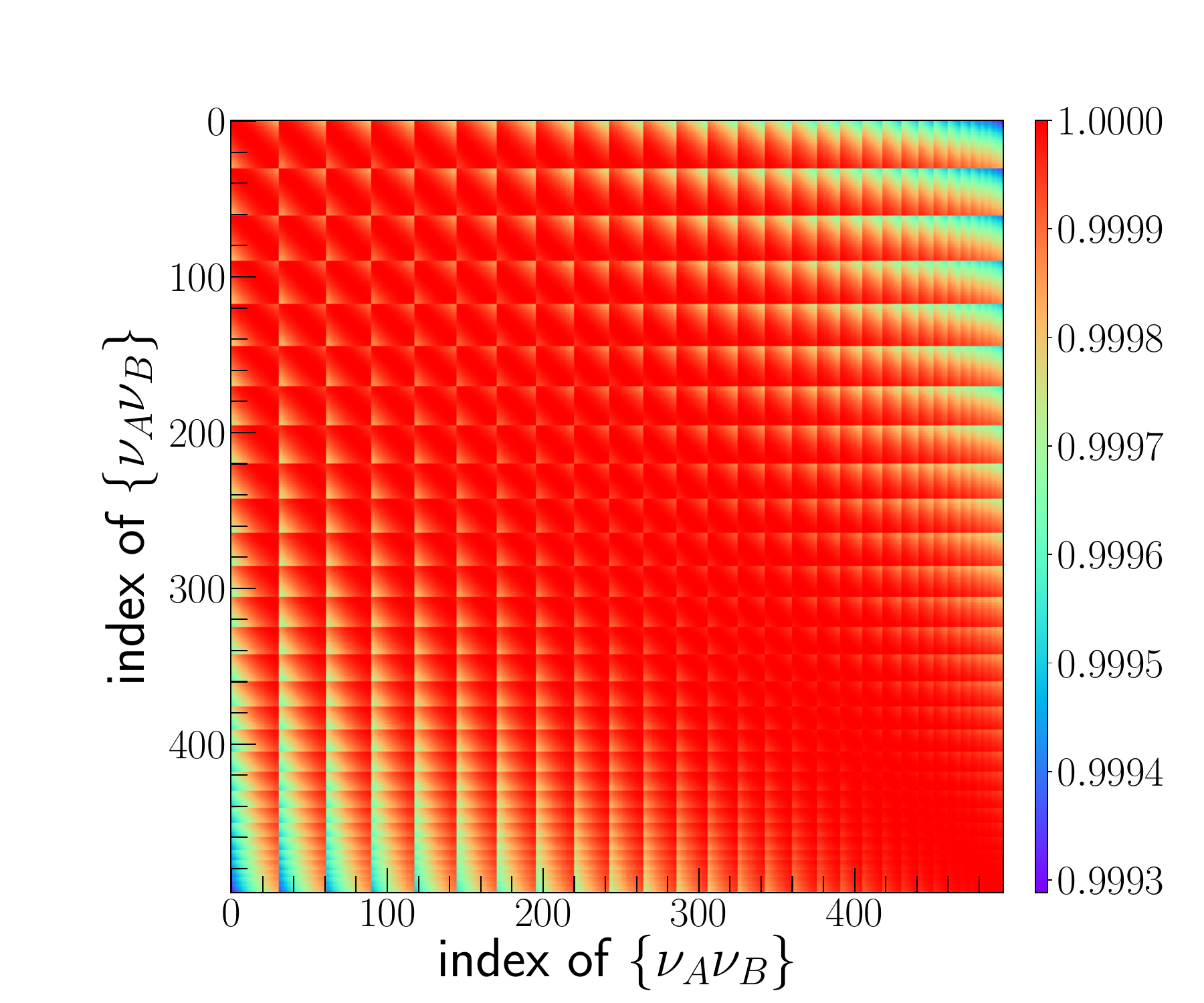}
\caption{Frequency correlation matrix for both neutral hydrogen signals (left) and the overall foreground contaminants (right) at the broadband $b$ = 162. The axis is arranged by the indices of all the frequency pairs $\{\nu_A\nu_B\}$. The foreground is strongly correlated among different frequency bands whereas the HI is only correlated at neighboring bands.}\label{corr}
\end{figure*}

Band-power correlations of mock data at different frequencies can be quantified by a covariance matrix 
\begin{equation}
\mathcal{C}_b\delta_{bb^{\prime}}=\langle({\bf C}_b-{\bf \bar C}_b)({\bf C}_{b^{\prime}}-{\bf \bar C}_{b^{\prime}})^T\rangle
\end{equation}
which can be analytically expressed by the Knox formula
\begin{eqnarray}
&&\mathcal{C}(C^{XY}_b,C^{WV}_{b^{\prime}})=\frac{1}{(2b+1)f_{\rm sky}\Delta b}(C^{XW}_bC^{YV}_b+C^{XV}_bC^{YW}_b)\delta_{bb^{\prime}}.\nonumber\\\label{knox}
\end{eqnarray}
Here a power spectrum vector is ${\bf C}=[C^{\nu_A\times\nu_B}_{b},C^{\nu_A\times\nu_B}_{b},C^{\nu_A\times\nu_B}_{b},...]^T$, $\{X, Y, W, V\}$ denotes a specific frequency, the $f_{\rm sky}$ is the sky fraction, $b$ is the averaged band center at $\ell<b<\ell^{\prime}$ and $\Delta b$ is the bandwidth of the angular power spectrum. ${\bf \bar C}_b$ is the averaged power-spectrum vector at the broadband $b$, $\langle...\rangle$ denotes ensemble average and the superscript ``$T$'' refers to the transpose of the vector.

From the map realization of different components in Fig. \ref{mock}, we can calculate the power spectra $C^{\nu_A\nu_B}_{\ell}$ at different frequency pairs $\{\nu_A\nu_B\}$. We arrange these power spectra in a two-dimensional space $\ell\mbox{-}\{\nu_A\nu_B\}$ for both foreground components and the HI in Fig. \ref{psdftRAW} and bin all the angular power spectra at $50<\ell<500$ with 6 broad bands. A representative plot for one-broadband band powers $C^{\{\nu_A\nu_B\}}_{b}$ at all frequencies is shown in Fig. \ref{psdft} and the derived correlation matrices for the foreground contaminants and the HI signals show different correlation patterns in the frequency domain as seen from Fig. \ref{corr}. 

From a theoretical perspective, we can split the full power-spectrum space into the cross-power ($\times$) spectrum space and the auto-power ($\circ$) spectrum space by applying a frequency-correlation length $\Delta\nu<\Delta\nu_{\ast}$ MHz as discussed in the main text. This implies that the HI signal is completely null at the cross-power spectrum subspace and the foreground signatures can be exactly determined without compromising any signals. We demonstrate this power-spectrum decomposition in Fig. \ref{psdft}. However, the map-domain methods, such as the principal component analysis~\cite{2015MNRAS.454.3240B} and generalized needlet internal linear combination~\cite{2016MNRAS.456.2749O}, have to deal with intensity mapping signals and foreground contaminants simultaneously, resulting in an inevitable signal loss. 

\begin{figure*}
\includegraphics[width=8cm, height=6cm]{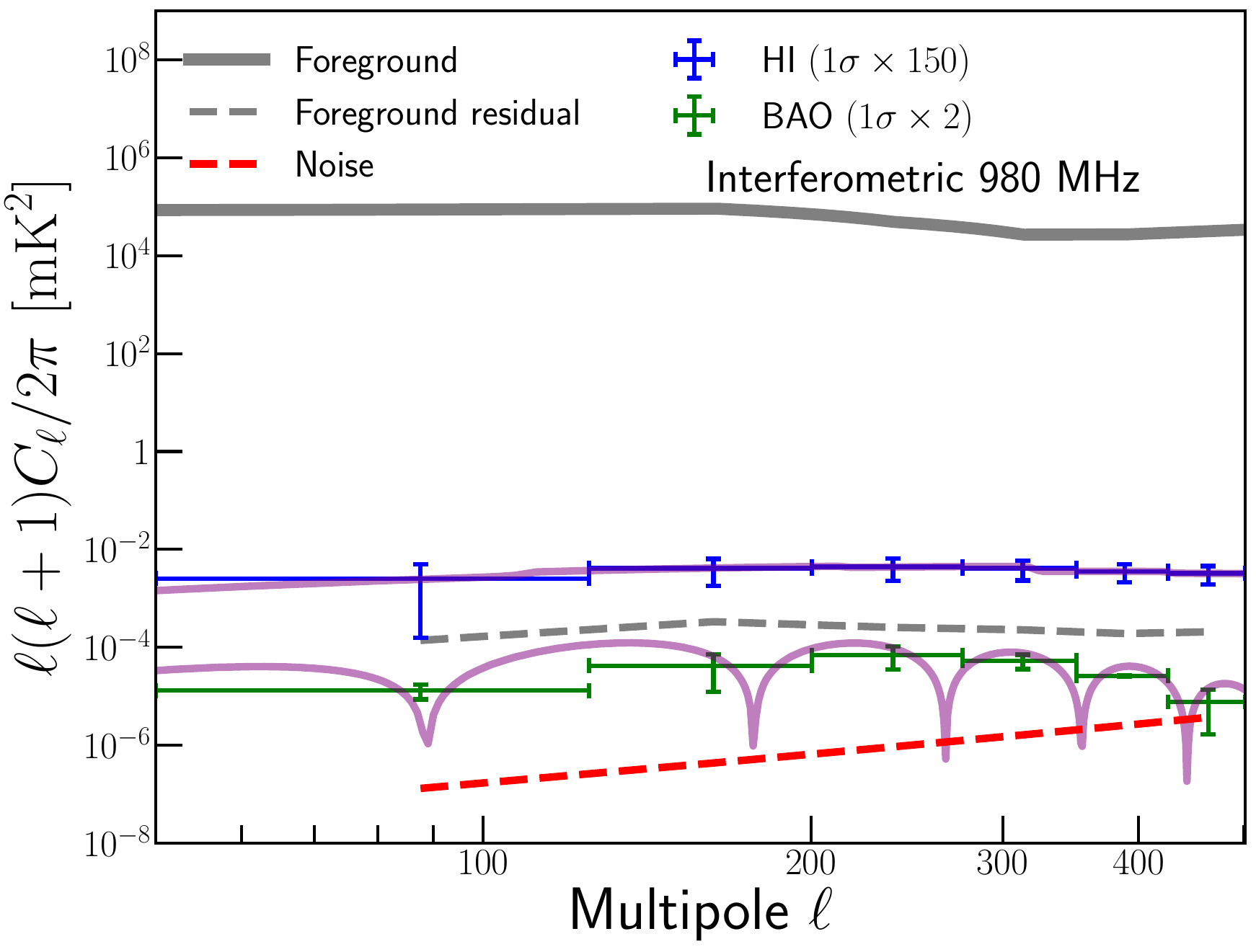}
\includegraphics[width=8cm, height=6cm]{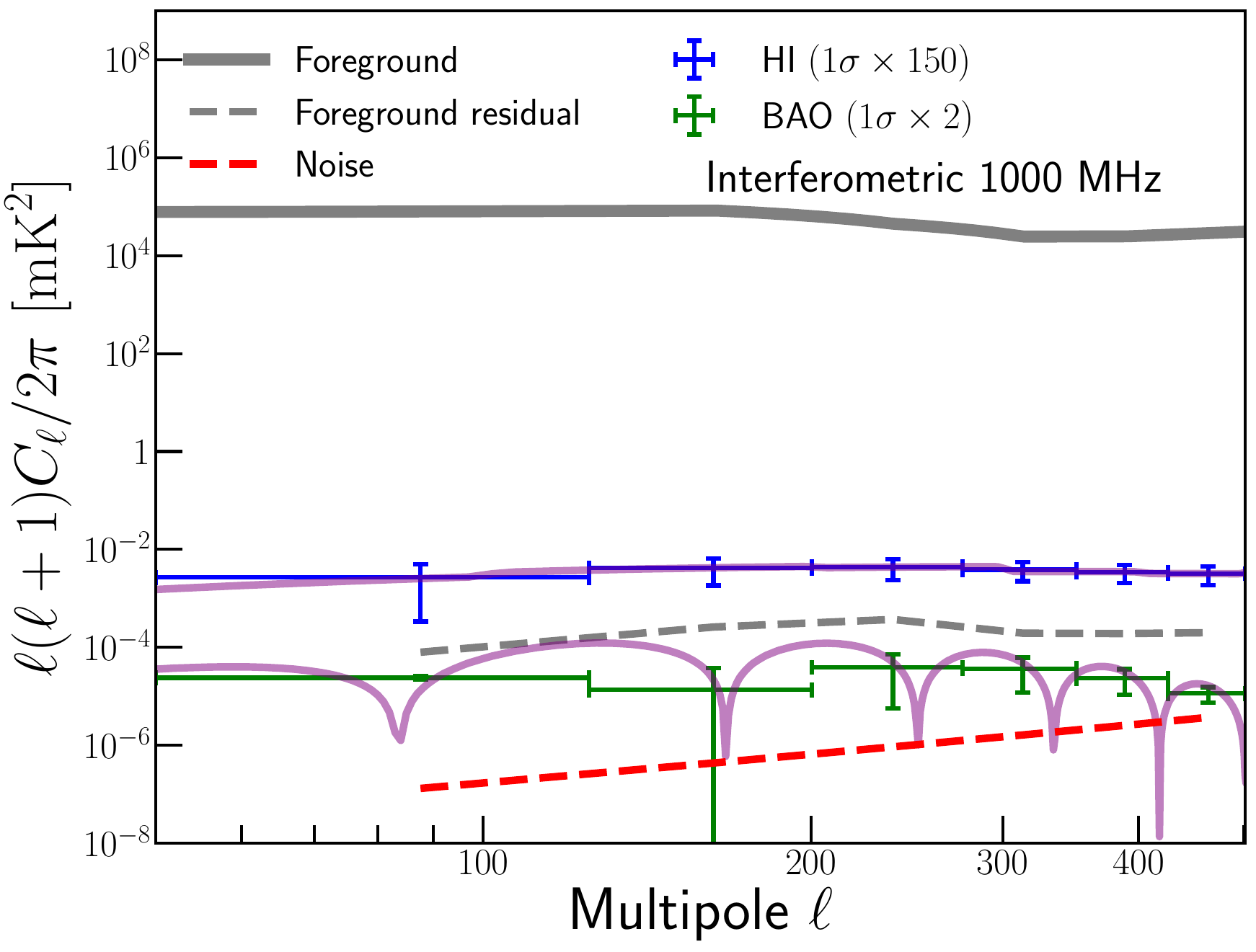}
\includegraphics[width=8cm, height=6cm]{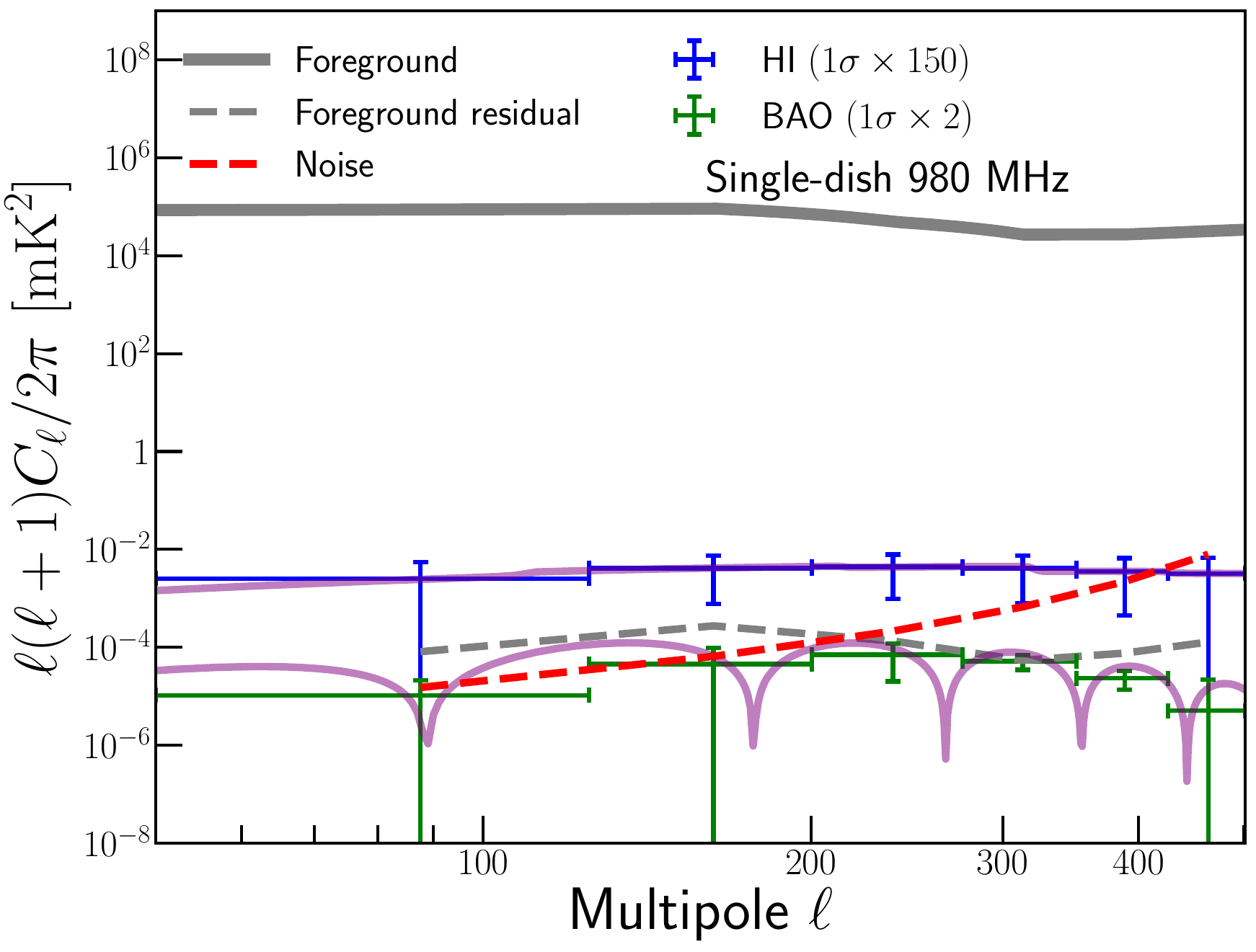}
\includegraphics[width=8cm, height=6cm]{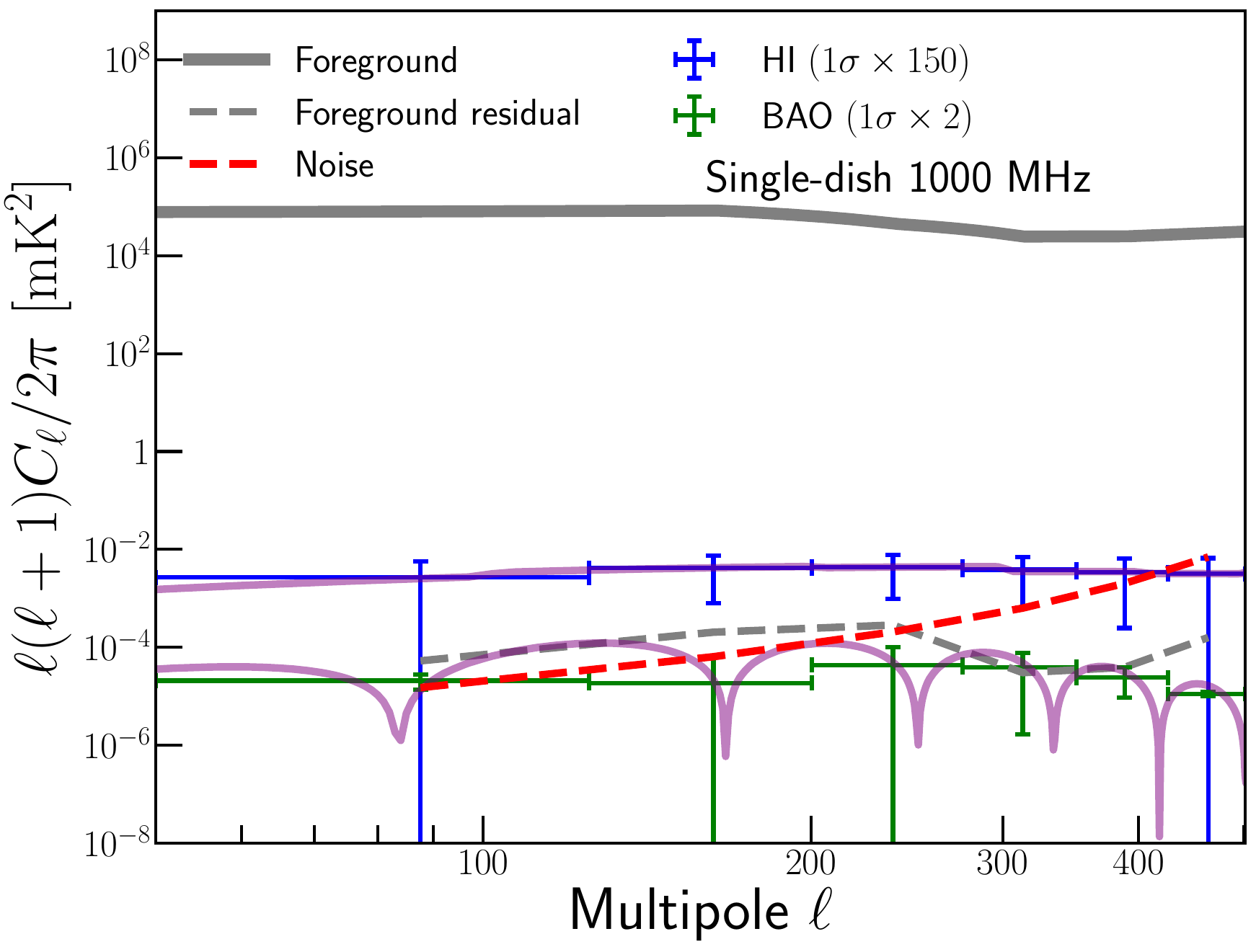}
\caption{Reconstructed band powers of foreground, HI (without BAO), and BAO at two representative frequencies (980, 1000 MHz) for both the interferometric (top) and single-dish-like IM experiments (bottom), respectively. Despite orders of magnitude differences in signal strengths, all the reconstructed band powers are consistent with theoretical expectations.}\label{bp}
\end{figure*}

We sample the parameter space using the Bayesian probability in Eq. (\ref{MLE}). The posterior distribution functions of the radio sky model are shown in Figs. \ref{pdf_ref} and \ref{pdf_dish} for both the interferometric and single-dish-like IM experiments, respectively. It is seen that all the parameters of the neutral hydrogen and BAO can be correctly reconstructed with high signal-to-noise ratios. The reconstructed band powers of foreground, HI, and BAO signals at two frequencies (980, 1000 MHz) are shown in Fig. \ref{bp} for both the interferometric (top) and single-dish-like IM experiments (bottom), respectively. All the reconstructed band powers are consistent with theoretical predictions.

\begin{figure*}
\includegraphics[width=16cm, height=16cm]{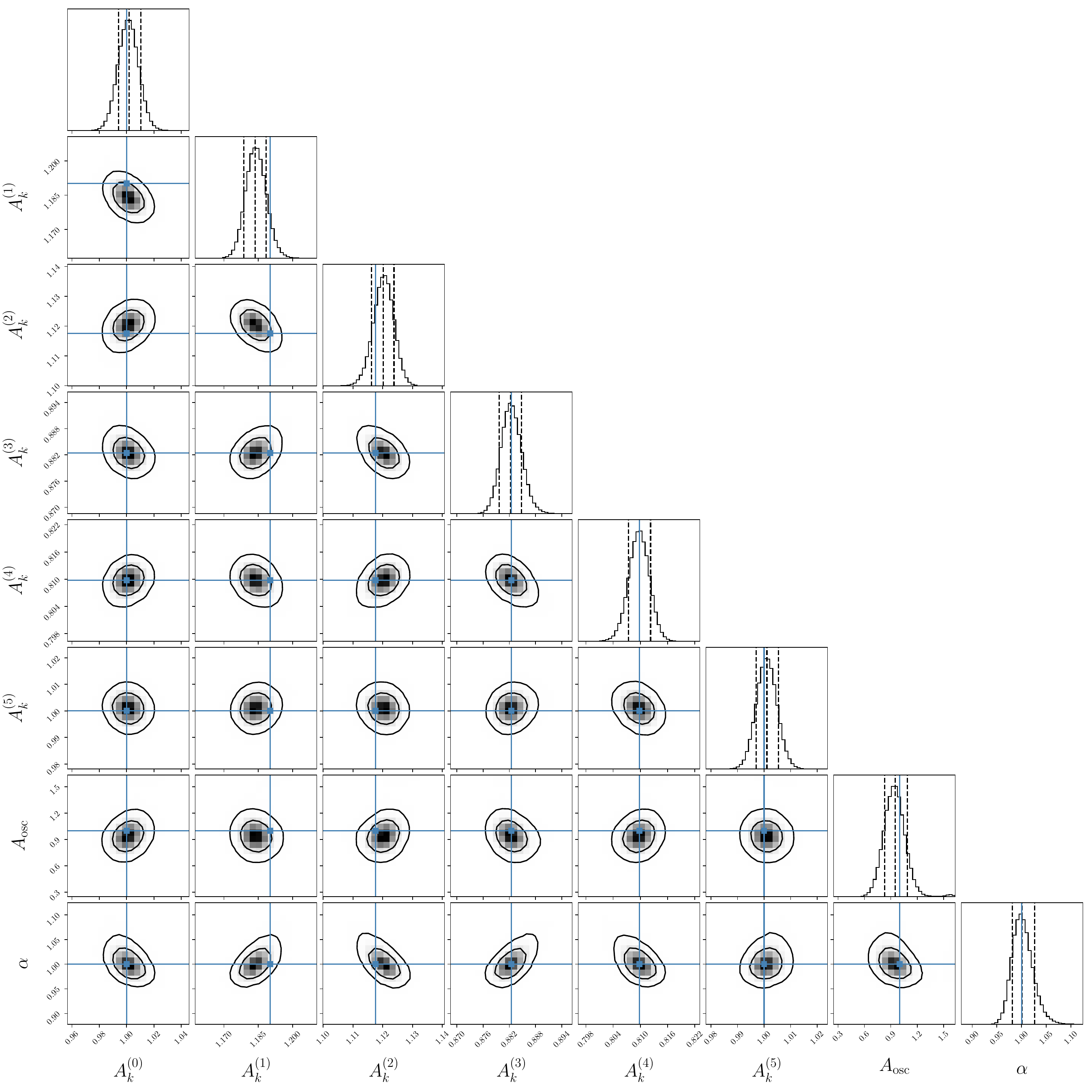}
\caption{Posterior distribution functions for the parameter set of radio sky model in Eq. (\ref{MLE}) for an interferometric IM experiment. All the parameters peak around the input values labeled by the vertical and horizontal lines, indicating that all the components are correctly reconstructed from the Bayesian analysis in Eq. (\ref{MLE}).}\label{pdf_ref}
\end{figure*}

\begin{figure*}
\includegraphics[width=16cm, height=16cm]{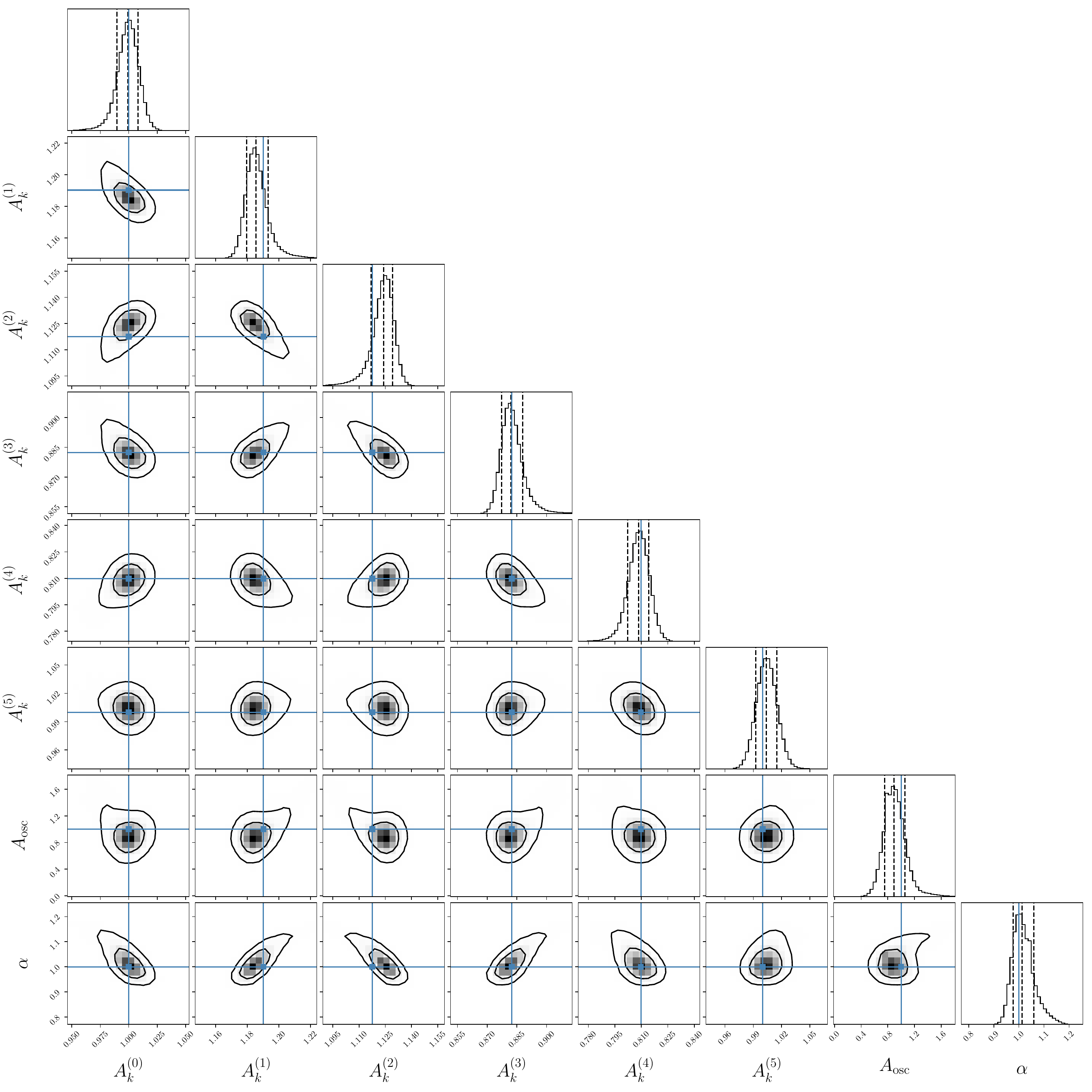}
\caption{Posterior distribution functions for the parameter set of radio sky model in Eq. (\ref{MLE}) for a single-dish-like IM experiment.}\label{pdf_dish}
\end{figure*}

\medskip

\end{document}